\documentclass[usegraphicx,usenatbib]{mn2e}
\usepackage{amssymb}
\usepackage{graphicx}
\usepackage{longtable}

\def\Teff{$T_{\rm eff}$}
\def\logg{$\log\,g$}

\def\Vt{V${\rm t}$}

\newcommand {\apgt} {\ {\raise-.5ex\hbox{$\buildrel>\over\sim$}}\ }
\newcommand {\aplt} {\ {\raise-.5ex\hbox{$\buildrel<\over\sim$}}\ }

\title[Molybdenum and Ruthenium]
{Enrichment of the Galactic disc with neutron-capture elements: Mo and Ru
}
\author[T.~Mishenina  et al.]
{T.~Mishenina$^{1}$ \thanks{tmishenina@ukr.net;* mpignatari@gmail.com},
 M.~Pignatari$^{2,3,6}$ \thanks{The NuGrid collaboration, http://www.nugridstars.org} *,
T.~Gorbaneva$^{1}$,  
C.~Travaglio$^{4,5}$ $\dagger$ ,\newauthor
B.~C\^{o}t\'e$^{3,6}$ $\dagger$ ,
F.-K.~Thielemann$^{7,8}$,
 C.~Soubiran$^{9}$  
 \\
$^{1}$Astronomical Observatory, Odessa National University,         
       Shevchenko Park, 65014, Odessa, Ukraine\\
$^{2}$ E.A. Milne Centre for Astrophysics, Dept of Physics \& Mathematics, University of Hull, HU6 7RX, United Kingdom\\
$^{3}$    Konkoly Observatory, Hungarian Academy of Sciences, Konkoly Thege Miklos ut 15-17, H-1121 Budapest, Hungary\\
$^{4}$ INFN, University of Turin, Via Pietro Giuria 1, 10025 Turin, Italy \\
$^{5}$ B2FH Association, Turin, Italy \\
$^{6}$ Joint Institute for Nuclear Astrophysics - Center for the Evolution of the Elements, USA\\
$^{7}$ Department of Physics, University of Basel, Klingelbergstrabe 82,
        4056 Basel, Switzerland\\
$^{8}$ GSI Helmholtzzentrum für Schwerionenforschung, Planckstrasse 1, D-64291 Darmstadt, Germany\\
$^{9}$  Laboratoire d'Astrophysique de Bordeaux, 
        Univ. Bordeaux  - CNRS, B18N,  all\'ee Geoffroy Saint-Hilaire, 33615 Pessac, France\\
}

\begin{document}

\date{Accepted 2015 xxx. Received 2015 xxx; in original form 2015 xxx}
\pagerange{\pageref{firstpage}--\pageref{lastpage}}
\pubyear{2015}

\maketitle

\label{firstpage}

\begin{abstract}
We present new observational data for the heavy elements molybdenum (Mo, Z = 42) and ruthenium (Ru, Z = 44) in F-, G-, and K-stars belonging to different substructures of the Milky Way. The range of metallicity covered is --1.0 $<$ [Fe/H] $<$ +0.3. The spectra of Galactic disc stars have a high  resolution of 42,000 and 75,000 and signal-to-noise ratio better than 100. Mo and Ru abundances were derived by comparing the observed and synthetic spectra in the region of Mo I lines at 5506, 5533 \AA~ for 209 stars and Ru I lines at 4080, 4584, 4757 \AA~  for 162 stars using the LTE approach. For all the stars, the Mo and Ru abundance determinations are obtained for the first time with an average error of 0.14 dex. This is the first extended sample of stellar observations for Mo and Ru in the Milky Way disc, and together with earlier observations in halo stars it is pivotal in providing a complete picture of the evolution of Mo and Ru across cosmic timescales.

The Mo and Ru abundances were compared with those of the neutron-capture elements (Sr, Y, Zr, Ba, Sm, Eu). The complex nucleosynthesis history of Mo and Ru is compared with different Galactic Chemical Evolution (GCE) simulations. 
In general, present theoretical GCE simulations show underproduction of Mo and Ru at all metallicities compared to observations. This highlights a significant contribution of nucleosynthesis processes not yet considered in our simulations. A number of possible scenarios are discussed.

\end{abstract}


\section{Introduction}

The elements molybdenum (Mo, Z = 42) and ruthenium (Ru, Z = 44) are located just above the first neutron-shell closure beyond iron, at N=50.
They both have seven stable isotopes, providing an ideal benchmark for nuclear astrophysics. The isotopes $^{92-94}$Mo and $^{94-96}$Ru are classically defined as p$-$only nuclei, i.e. they can be made by the $\gamma$ process or by some $p$--process component, but not by neutron-capture processes. Their high concentration in the solar system compared to other neutron-rich Mo and Ru isotopes is still a major puzzle to be solved \citep[e.g.][and references therein]{arnould:03,rauscher:13,pignatari:16a}. $^{96}$Mo and $^{100}$Ru are classified as $s$-only isotopes, i.e. they can be made by the slow neutron-capture process or $s$--process \citep[e.g.][and references therein]{kaeppeler:11}. $^{100}$Mo and $^{104}$Ru are not efficiently produced via the $s$-process \citep[e.g.][]{bisterzo:14}. They are classified as $r$-only isotopes, i.e. they are made mostly by some of the rapid neutron-capture process components, or $r$-process \citep[][and references therein]{cowan:19}.  
Finally, the intermediate neutron-capture process \citep[$i$-process;][]{cowan:77} have been shown to produce efficiently the Mo stable isotopes $^{95}\mathrm{Mo}$ and $^{97}\mathrm{Mo}$, and preliminary evaluations of the $i$-process contribution to Mo and Ru have been reported in \citep{cote:18a}.  
Therefore, studying these elements in the context of Galactic Chemical Evolution (GCE) can provide valuable diagnostics on the nucleosynthesis processes described above.

The $s$-process component of Mo and Ru is made by the main $s$-process component between Sr and Pb, that is produced by Asymptotic Giant Branch stars \citep[AGB stars, e.g.][]{gallino:98, busso:99}. 
In these stars most of the neutrons are released by the $^{13}\mathrm{C}$($\alpha$,n)$^{16}\mathrm{O}$ reaction in the radiative $^{13}\mathrm{C}$-pocket, formed right after the third dredge-up event \citep[e.g.][]{straniero:95}. The rest of the neutrons are supplied by the  
partial activation of the $^{22}\mathrm{Ne}$($\alpha$,n)$^{25}\mathrm{Mg}$ reaction during the convective thermal pulse \citep[][and reference therein]{busso:99,herwig:05,karakas:14}.  
The weak $s$-process component in the solar system originates in massive stars, and is mostly due to the $^{22}\mathrm{Ne}$($\alpha$,n)$^{25}\mathrm{Mg}$ activation in the convective He-burning core and in the convective C-burning shell \citep[e.g.][and references therein]{the:07, pignatari:10}. The weak $s$-process contributes to no more than few per cent to the solar abundance of Mo and Ru \citep[][]{travaglio:04}. The relevance of additional $s$-process production in rotating massive stars to GCE is still being debated, in the early Galaxy as well as for the solar system \citep[e.g][]{pignatari:08,maeder:15,cescutti:15,frischknecht:16,choplin:18}. 

The origin of the $r$-process abundances beyond Fe is still matter of debate. Several 
astrophysical scenarios have been proposed: 
1) neutrino-driven winds from core-collapse supernovae (CCSNe) \citep[e.g.][]{hoffman:94, hoffman:97, wanajo:01, farouqi:09, arcones:11, kratz:14} or electron-capture supernovae (ECSNe), i.e. collapsing O-Mg-Ne cores \citep{wanajo:11} (weak $r$-process); 2)  neutron-rich matter ejected by neutron star mergers \citep[e.g.][]{freiburghaus:99, goriely:11, wu:16} and neutron star - black hole mergers \citep{surman:08, wehmeyer:19} (main $r$-process);
3) ejecta from rotating MHD core-collapse supernovae and/or collapsars  \citep[e.g.][]{nishimura:06,winteler:12,nishimura:17,mosta:18,siegel:19}. The origin of $r$-process elements in the Milky Way has been discussed recently by \cite{cote:18b} and reviewed by \cite{cowan:19}. 

The (classical) $p$-process is identified with explosive Ne/O-burning in outer zones of the progenitor star. It is initiated by the passage of the supernova shock wave and acts via photodisintegration reactions which produce neighboring (proton-rich) isotopes from pre-existing heavy nuclei \citep[see][and references therein]{arnould:03, rauscher:13, pignatari:16a}. 
The most established scenario proposed for the $p$-process production are
Type~II supernova explosions \citep[][]{woosley:78,rayet:95}, with a potential relevant contribution from the advanced pre-supernova stages \citep{arnould:76, rauscher:02, ritter:18}.  
Complementary scenarios are Type Ia Supernovae \citep[][]{howard:91, travaglio:11, travaglio:15,nishimura:18} and He-accreting CO white dwarfs of sub-Chandrasekhar mass \cite[][]{goriely:02}. 
Proton-rich components of neutrino-driven winds have also been proposed as a potential relevant source for the light $p$-process nuclei \citep[e.g.][]{frohlich:06,frohlich:17,martinez:14,eichler:18}, although their effective contribution in the Mo and Ru region is challenged by observations of radioactive $^{92}$Nb abundance in the early solar system \citep[][]{dauphas:03}.
We refer to \cite{pignatari:18}, \cite{wanajo:18} and \cite{bliss:18} for the most up-to-date theoretical data on the production of Mo and Ru in CCSNe.
 
The solar abundances of Mo and Ru isotopes adopted from \cite{anders:89} are listed in the second column of Table~\ref{diff_sun}. The most important isotopes contributing to the Mo and Ru abundances are $^{98}\mathrm{Mo}$ and $^{102}\mathrm{Ru}$.
As we mentioned before, the isotopes $^{92, 94}\mathrm{Mo}$ isotopes are produced by the $p$-process.
The table also shows the $s$-process contribution to Mo isotopes derived by \cite{travaglio:04} using GCE simulations and stellar yields for low- and intermediate-mass stars (LIM), as well as the $s$-process contribution to the solar composition estimated using the GCE simulations for Mo and Ru by \cite{arlandini:99}, \cite{travaglio:04} and \cite{bisterzo:14}.

\begin{table*}
\caption{Contribution of Mo and Ru isotopes to the solar abundances: 1 - Anders\&Grevess (1989),  2 - Arlandini et al. (1999); 3 - Travaglio et al.(2004); 4 - Bisterzo et al. (2014). }
\label{diff_sun}
\begin{tabular}{lccccc}
\hline
\hline
ELEMENT   &     SOLAR (\%) &  $s$-process (no GCE) & $s$-process + GCE (\% ) &  $s$-process + GCE (\%) & $p$-process (\%)    \\
\hline
       &   1  &  2  & 3  &  4   &    \\
\hline
$^{92}\mathrm{Mo}$     &        14.84&     &                   &      &     100   \\
$^{94}\mathrm{Mo}$     &         9.25&     &                   &      &        100 \\
 $^{95}\mathrm{Mo}$    &         15.92&     &                39 &      &             \\     
 $^{96}\mathrm{Mo}$    &         16.68&    &                78 &      &            \\
 $^{97}\mathrm{Mo}$    &         9.55 &    &                46 &      &            \\
 $^{98}\mathrm{Mo}$    &         24.13&    &                 59 &      &            \\
 $^{100}\mathrm{Mo}$   &         9.63  &   &                    &      &                   \\
\hline
Mo                     &               & 50  &             38    &   39    &             \\
\hline 
$^{96}\mathrm{Ru}$     &          5.52&   &                   &       &    100       \\
$^{98}\mathrm{Ru}$     &          1.88&   &                   &       &       100       \\
$^{99}\mathrm{Ru}$     &          12.7&   &                   &       &                  \\
$^{100}\mathrm{Ru}$    &          12.6&   &                   &       &                      \\
$^{101}\mathrm{Ru}$    &          17.0&   &                   &       &                      \\
$^{102}\mathrm{Ru}$    &         31.6&   &                   &       &                      \\
$^{104}\mathrm{Ru}$    &          18.7&    &                   &       &                      \\
\hline
Ru                     &              & 32 &            24        &    29   &                    \\
\hline 
\hline
\end{tabular}
\end{table*}

This paper is the last one in a series of those focused on the observations of different elements in the Galactic disc. In the first studies, particular attention was paid to the enrichment of the thin and thick disc stars with the $\alpha$-elements and neutron-capture elements \citep{mishenina:04, mishenina:13}, as well as Mn \citep{mishenina:15}, and Sr \citep{mishenina:19}. Stellar observations for our sample of stars and different data sets have been compared with a number of GCE simulations \citep{mishenina:17}. In this work, we focus on Mo and Ru. 
Although these elements have been investigated in metal-poor stars \citep[e.g.][]{ivans:06, peterson:11, peterson:13, roederer:12, hansen:14, sakari:18}, there is a lack of observations at higher metallicities ([Fe/H]) between $-0.7$ and 0.3, which is the range covered in this study.
We aim at providing the first extended sample of stellar observations for Mo and Ru abundances in Galactic disc stars and analyzing their chemical signatures using theoretical GCE models.

The paper is structured as follows. 
The observations and selection of stars along with the definition of the main stellar parameters are described in \S \ref{sec: stellar param}. 
 \S \ref{sec: abundance determination} presents the abundance determinations and analysis of corresponding errors. 
The application of the results in the theory of nucleosynthesis and the chemical evolution of the Galaxy is discussed in \S \ref{sec: result, gce}. 
And finally, \S \ref{sec: conclusions} summarizes the finding and presents the conclusions drawn.

\section{Observations and atmospheric parameters}
\label{sec: stellar param}
In this investigation, we used the same spectra, atmospheric parameters and analytical techniques as earlier in \citep{mishenina:13}. 
The spectra of the target stars were obtained using the 1.93 m telescope at Observatoire de Haute-Provence (OHP, France) equipped with the echelle-type spectrograph ELODIE \citep{baranne:96} with the resolving power of R\,=\,42,000, the wavelength range from 4400 to 6800 \AA and signal to noise (S/N) ratio of about 100 -- 300. We also used additional spectra taken from the OHP spectroscopic archive  \citep{moultaka:04}, presenting the SOPHIE spectrograph \citep{perruchot:08} data covering a similar wavelength range at the resolution of R\,=\,75,000.

The online initial processing of spectra was carried out during observations \citep{kratz:98}. Further spectra processing such as the continuum arrangement, and measurements of the line depths and equivalent widths (EW), was conducted using the DECH30 software package developed by G.A. Galazutdinov (2007), http://gazinur.com/DECH-software.html.

The stellar atmospheric parameters of our target stars were determined earlier using uniform techniques for all the studied stars. The procedures employed to derive the effective temperatures \Teff\, surface gravities \logg, and microturbulent velocity  \Vt\ for our stars were described in detail in \cite{mishenina:01} and \cite{mishenina:04,mishenina:08}. 
The effective temperatures \Teff\ were derived by the calibration of the line-depth ratios for spectral line pairs that have  different low-level excitation potentials \citep{kovtyukh:03}.  
For the most metal-poor stars in the sample, \Teff\ were estimated by adjusting the far-wings of the H$_\alpha$ line \citep{mishenina:01}.
The surface gravities \logg\ were computed by the ionization balance, implying that similar iron abundances were obtained from the neutral iron Fe~{\sc i} and ionised iron Fe~{\sc ii} lines. 
The microturbulent velocity \Vt\ was established by factoring out the correlation  between the abundances and the equivalent widths of the Fe~{\sc i} lines.
We used the Fe~{\sc i} lines to derive the metallicity [Fe/H].

We compared our atmospheric parameters with the results of other authors in \cite{mishenina:04, mishenina:08, mishenina:13, mishenina:19}.
The estimated accuracy of our parameter determinations is as follows: 
$\Delta$\Teff\,=\,$\pm100~K$, surface gravities $\Delta$\logg\,=\,$\pm0.2$dex and microturbulent velocity 
$\Delta$\Vt\,=\,$\pm0.2$km s$^{-1}$. 

In this paper, we also compare our parameter determinations with those obtained in other studies for the stars common to our sample \citep{delgado:17, battistini:16, adibekyan:14, nissen:11, feltzing:07, takeda:07, brewer:06, reddy:03, mashonkina:01}. The mean differences between the parameters,  the errors and the number of common stars are given in Table \ref{ncap}.  In general, we see a good agreement of our findings with the results of other authors.

\begin{table*}
\begin{center}
\caption[]{Comparison of obtained stellar parameters with those reported by other authors for the n stars common to our sample. The full list of stellar parameters applied for each star in this study is provided in Table \ref{comp_edd} and in Mishenina et al. (2019).}
\label{ncap}
\begin{tabular}{lccccc}
\hline
\hline
 Reference & $\Delta$(\Teff, K) & $\Delta$(\logg) & $\Delta$([Fe/H]) &  n \\
\hline
Delgado Mena et al. (2017)& 27$\pm$36 & -0.08$\pm$0.13 & -0.01$\pm$0.03 &  12 \\
Battistini \& Bensby (2016)& -4$\pm$106 & -0.10$\pm$0.15 & -0.03$\pm$0.06 &  22 \\
Adibekyan et al. (2014)& 28$\pm$57 & -0.07$\pm$0.14 & 0.01$\pm$0.04 &  9  \\
Nissen et al. (2011)& 7$\pm$143 & -0.03$\pm$0.20 & -0.05$\pm$0.10  & 4 \\
Feltzing et al. (2007)& 24$\pm$76 & -0.03$\pm$0.13 & -0.01$\pm$0.08 &  10 \\
Takeda et al. (2007)& -14$\pm$119 & -0.06$\pm$0.21 & -0.04$\pm$0.10 &  31  \\
Brewer \& Carney (2006) & 64$\pm$112 & 0.02$\pm$0.20 & 0.09$\pm$0.06 &  4 \\
Reddy et al. (2003)& 127$\pm$13 & -0.08$\pm$0.14 & 0.09$\pm$0.02 &  7  \\
Mashonkina \& Gehren (2001) & 26$\pm$56 & -0.10$\pm$0.21 & 0.03$\pm$0.06 &  14  \\
\hline
\hline
\end{tabular}
\end{center}
\end{table*}

We adopt the kinematic classification of the stars into the thin and thick discs and Hercules stream, as described in \cite{mishenina:13}. We have not updated our classification with respect to the latest astrometric data from the Gaia Data Release 2 \citep{GDR2:18} because the stars in our sample are bright and tend to have Gaia astrometric errors equivalent to those of the Hipparcos observations. Some stars are even too bright to be measured by Gaia.
Our previous sample (276 stars), contained 21 stars belonging to the thick disc, 212 to the thin disc, 16 to the Hercules stream, and 27 are unclassiﬁed.

\section{Determination of Mo and Ru abundances}
\label{sec: abundance determination}

\begin{figure}
\begin{tabular}{c}
\includegraphics[width=8cm]{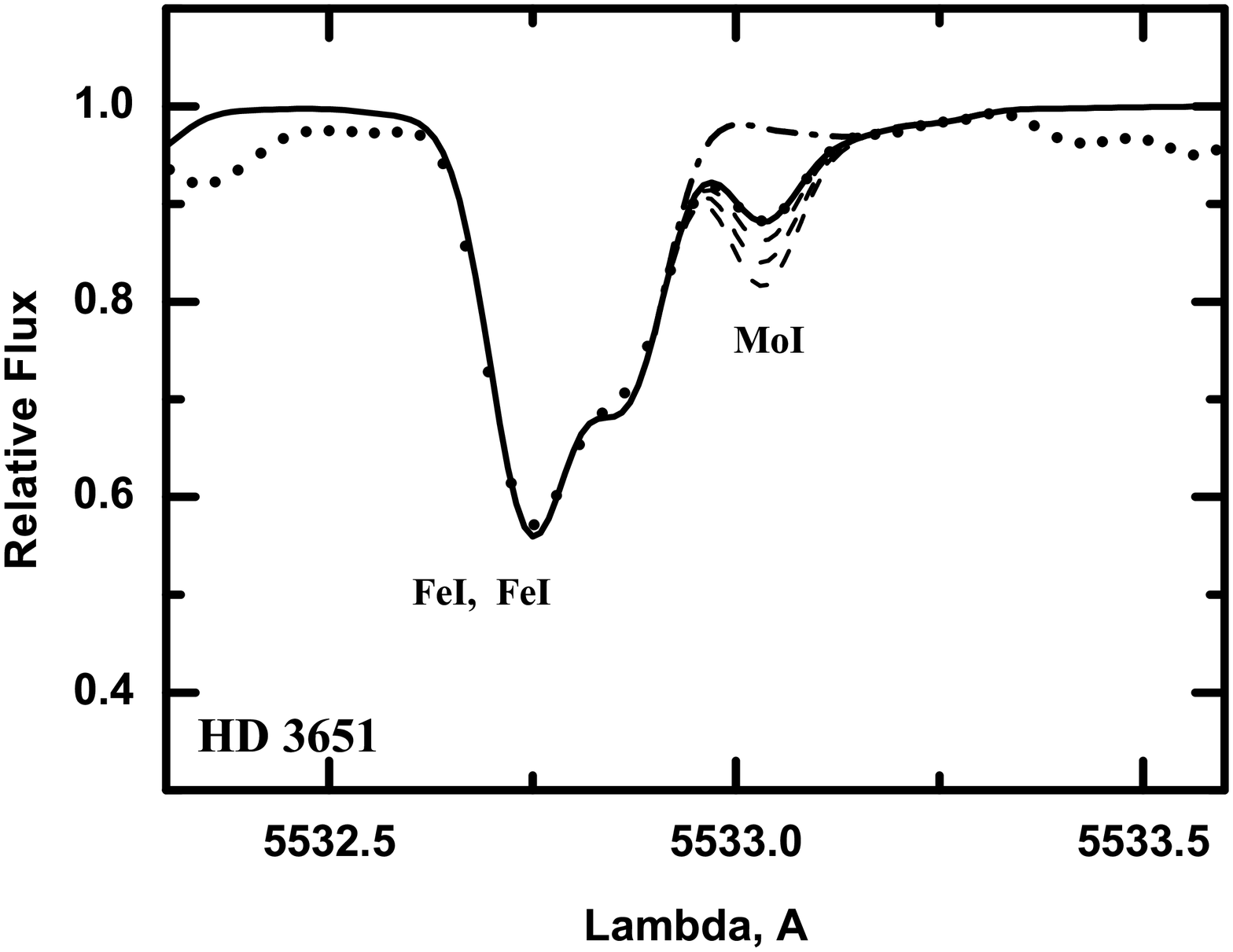}\\
\includegraphics[width=8cm]{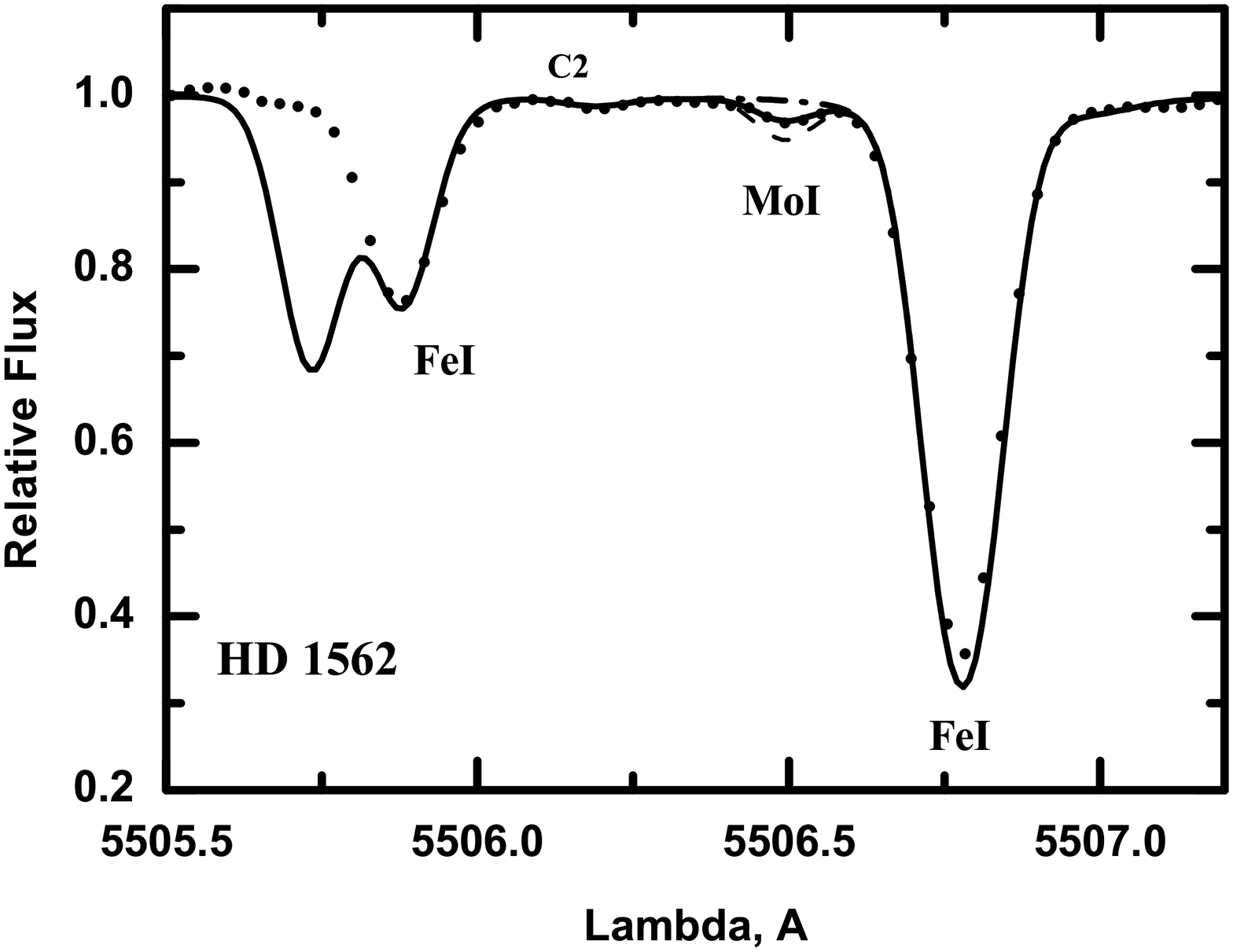}\\
\includegraphics[width=8cm]{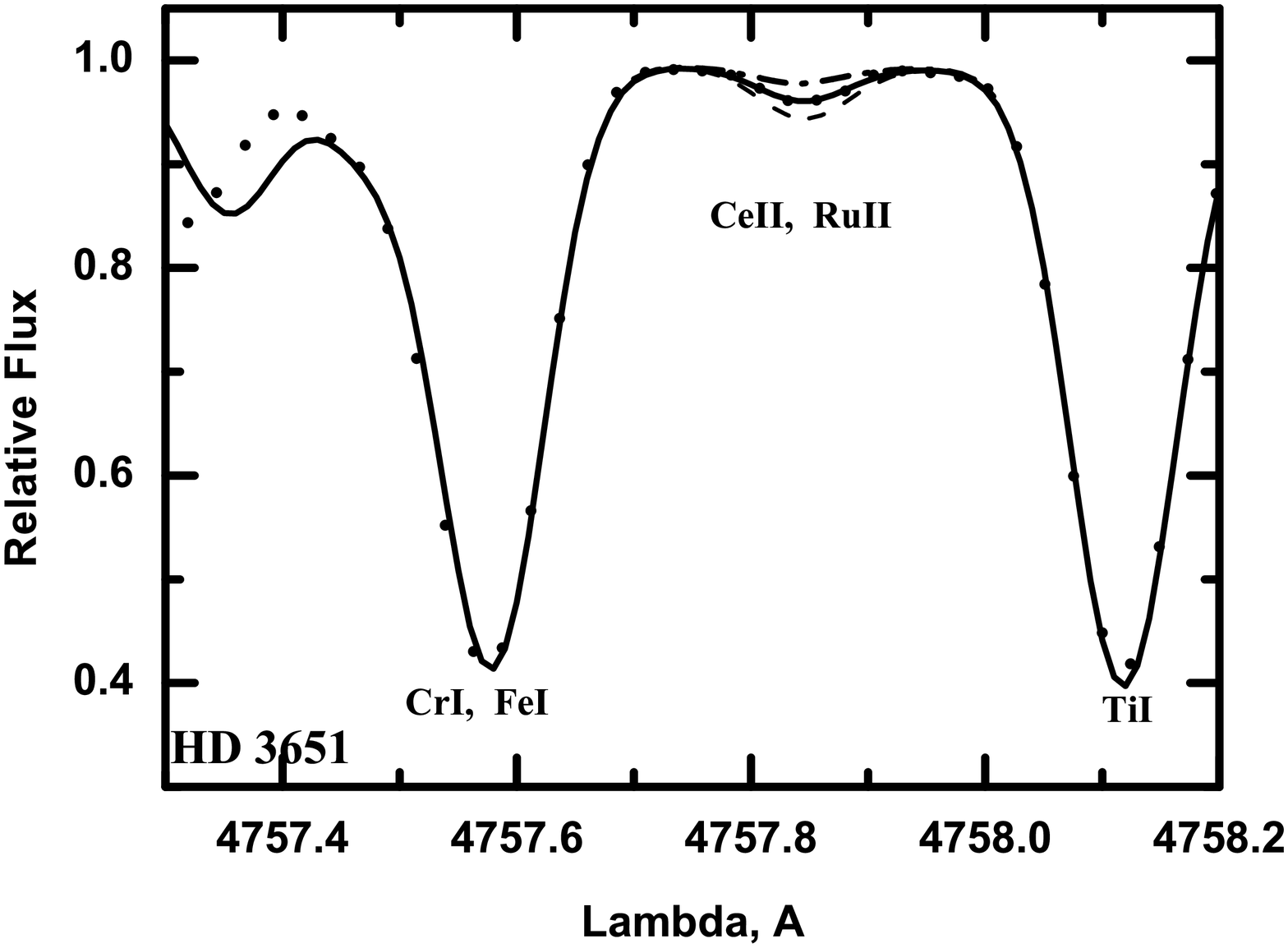}\\
\includegraphics[width=8cm]{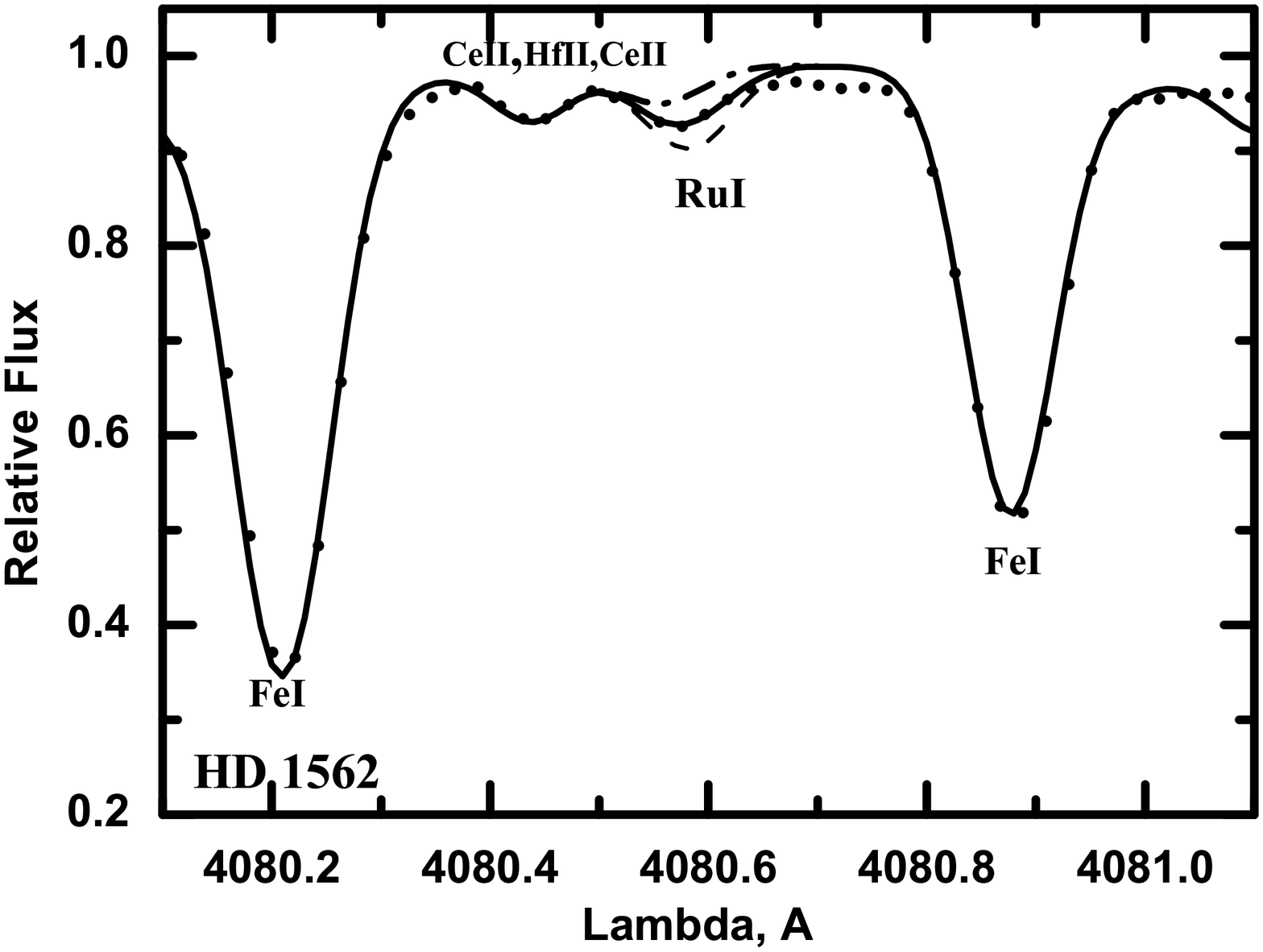}\\
\end{tabular}
\caption{Observed (dotted) and calculated (solid)  spectra in the region of Mo I and Ru I lines for the stars HD 1562 and HD 3651. The dashed lines correspond to the difference in 0.1 dex for the Mo abundance in HD 3651 and 0.3 dex in other cases.}
\label{moru_prof}
\end{figure}

The Mo and Ru abundances were derived using the LTE approximation applying the models of \cite{castelli:04} and the modified STARSP LTE spectral synthesis code \citep{tsymbal:96}. For Mo I lines at 5506, 5533 \AA~, and Ru I lines at 4080, 4584, and 4757 \AA~, the oscillator strengths log\,gf were adopted from last version (2016) of the VALD database \citep{kupka:99}. 
Both Mo I lines are fairly well measured in the spectra of our target stars. The Mo I 5533 \AA~ line is represented in the list of the Gaia-ESO Survey (GES), and and has been used by \cite{hansen:14} for the investigation of 71 meta-poor stars.
The comparison of synthetic and observed spectra for the Mo I and Ru I lines is shown in Fig. \ref{moru_prof}. 

The adopted LTE solar Mo and Ru abundances are log A(Mo)$_\odot$ = 1.88$\pm$0.08 and log A(Ru)$_\odot$ = 1.75$\pm$0.08 \citep{asplund:09}. 

\par
\begin{figure}
\begin{tabular}{c}
\includegraphics[width=8cm]{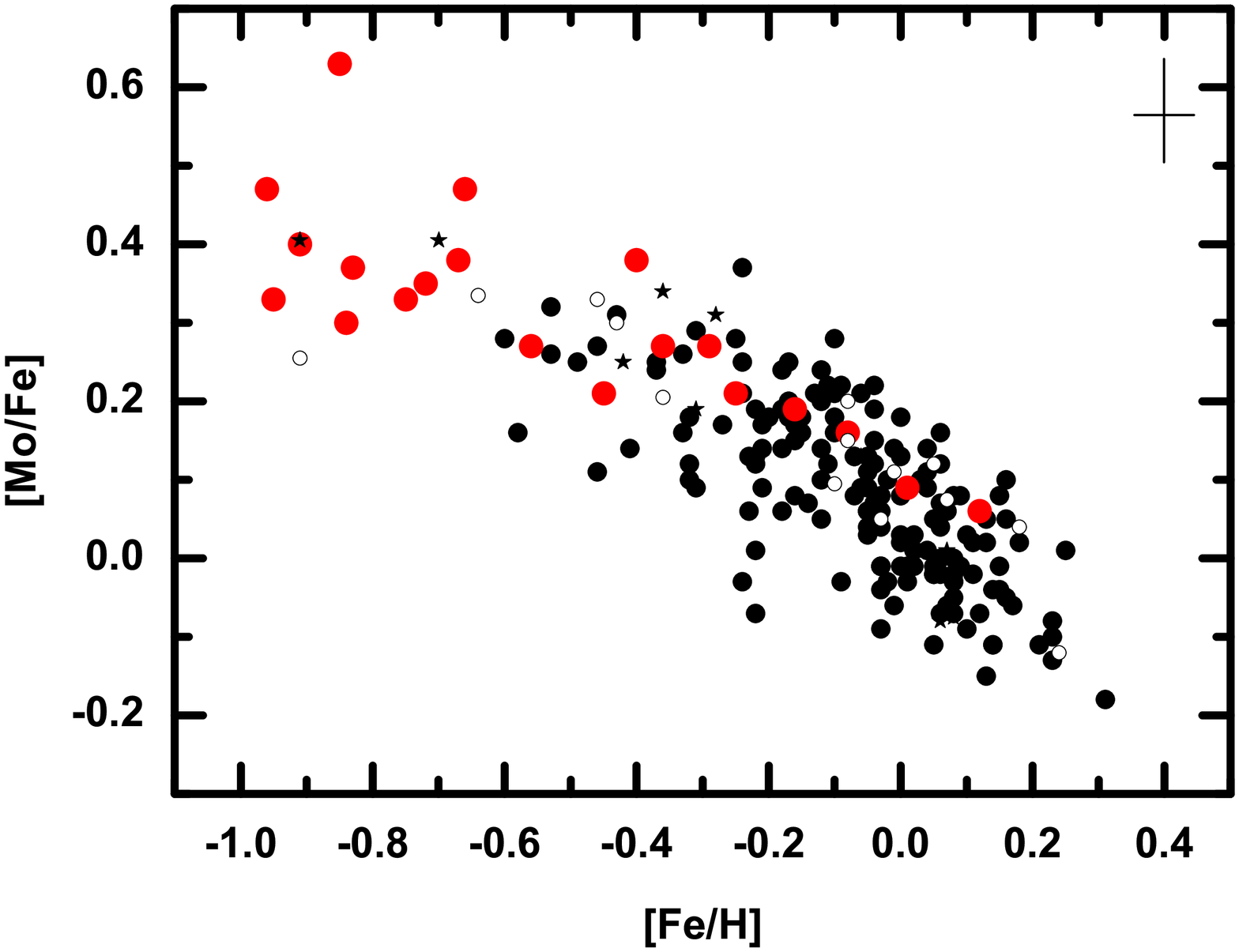}\\
\includegraphics[width=8cm]{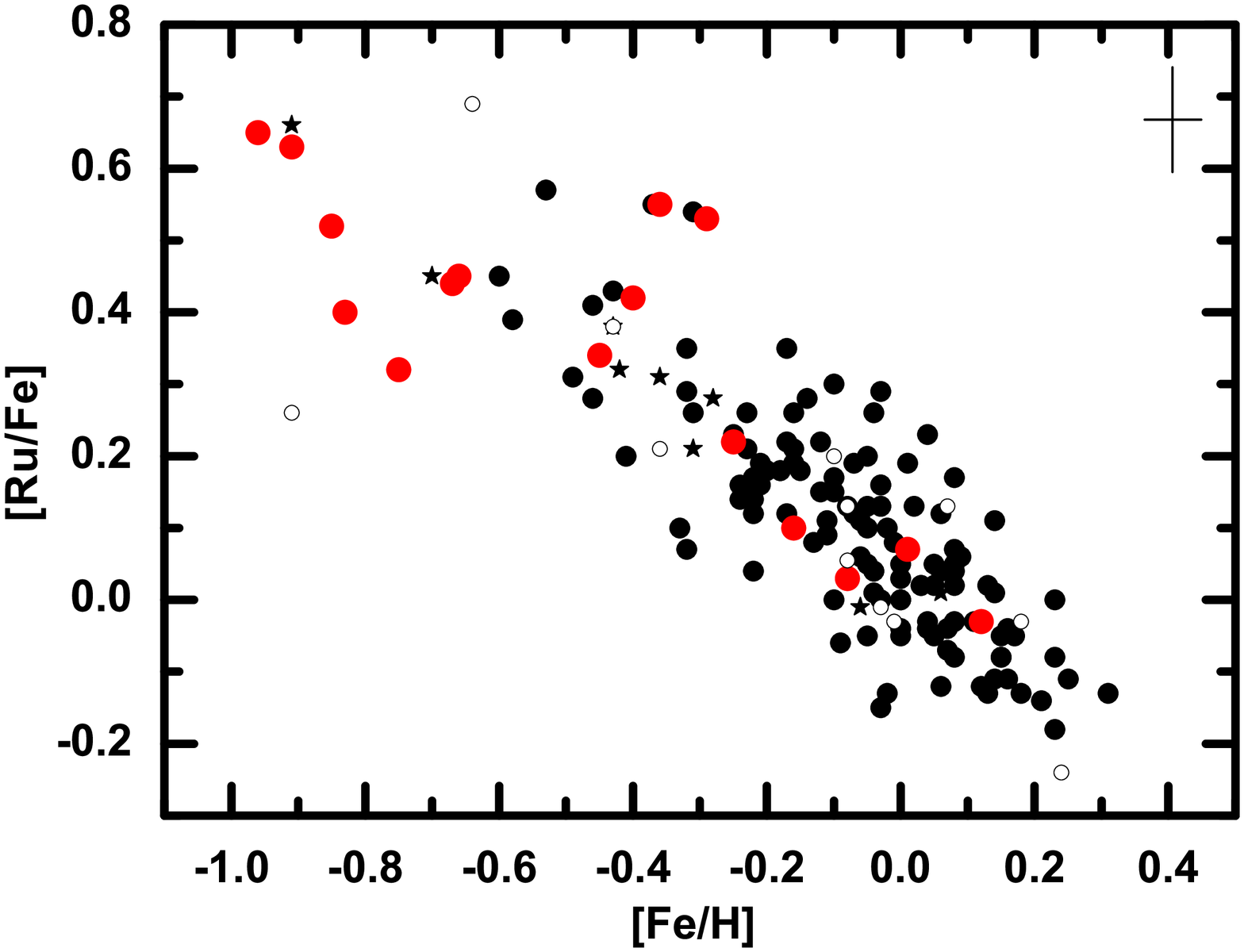}\\
\end{tabular}
\caption{[Mo/Fe] and [Ru/Fe] as a function of [Fe/H]. The stars belonging to the thin and thick discs are marked with small black and red circles, respectively. The stars classified into the Hercules stream are marked with asterisks while non-classified stars are depicted as open circles. The error bar is marked with a cross in the upper-right corner. }
\label{mo_ru_fe}
\end{figure}

We determined Mo abundance for 163 stars of the thin disc, 20 stars of the thick disc, 12 stars in the Hercules group, and 14 unclassified stars, which represents a total of 209 stars. Accordingly, the Ru content was determined for 124, 16, 10 and 12 stars belonging to the considered substructures, which made 162 stars in total.
The obtained Mo and Ru abundances, as well as stellar parameters, are given in Table A1.
Fig. \ref{mo_ru_fe} shows our [Mo/Fe] and [Ru/Fe] data as a function of [Fe/H].

\subsection{Errors in abundance determinations}

\begin{figure}
\begin{tabular}{c}
\includegraphics[width=8cm]{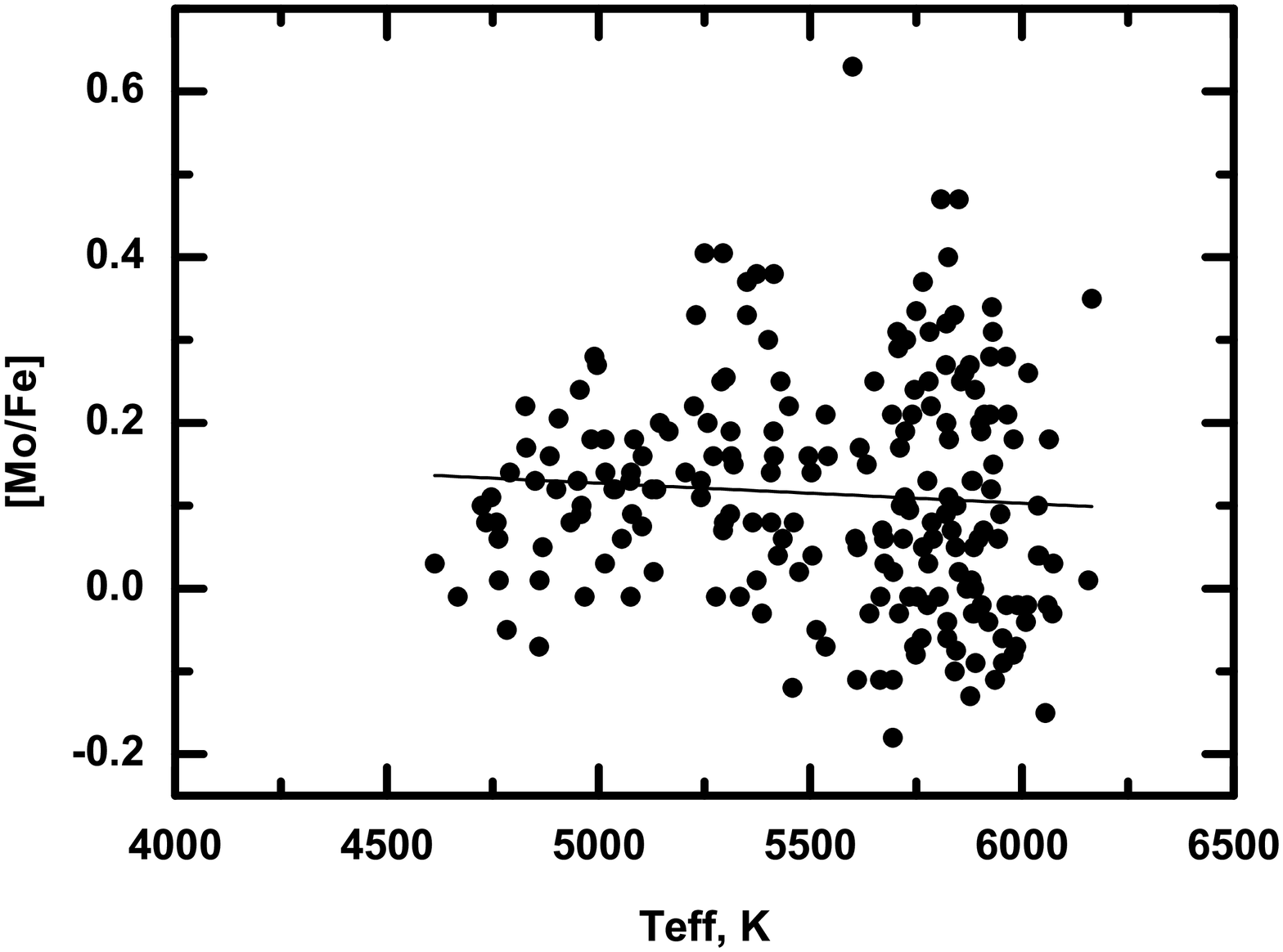}\\
\end{tabular}
\caption{Dependence of [Mo/Fe] on \Teff.}
\label{mo_teff_}
\end{figure}

\begin{figure}
\begin{tabular}{c}
\includegraphics[width=8cm]{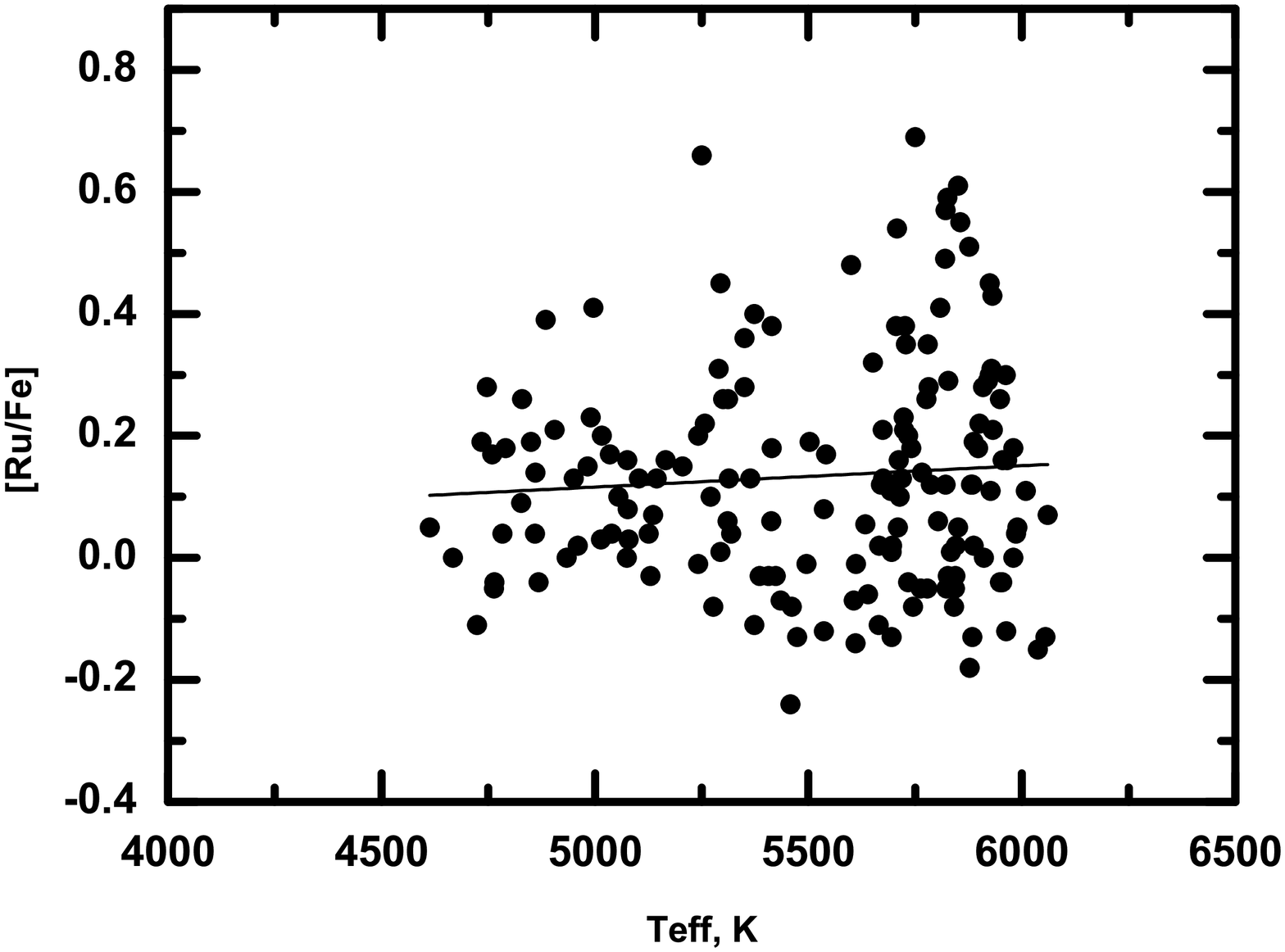}\\
\end{tabular}
\caption{Dependence of [Ru/Fe] on \Teff.}
\label{ru_teff_}
\end{figure}

We estimated systematic errors in the abundance of molybdenum and ruthenium abundance determinations due to the uncertainty of the atmospheric parameters on the basis of the results obtained for two stars - namely, HD154345 (\Teff\ = 5503 K; \logg = 4.30; \Vt = 1.3 km s$^{-1}$; [Fe/H ] = -0.22) and HD82106 (\Teff\ = 4827 K; \logg = 4.10; \Vt = 1.1 km s$^{-1}$; [Fe/H] = -0.11); we  used the Mo, Ru abundances for several models with modified parameters
($\Delta$ \Teff$ = \pm100 ~ K $, $\Delta$ \logg $ = \pm0.2$, $\Delta$ \Vt $ = \pm0.1 $).
The obtained variations of the abundance for different parameters and the adjustment errors for the calculated and observed spectral line profiles (0.02 dex) are given in the table \ref{errors}. The error in the \Teff determination is the major contributor to the error in the Mo and Ru abundance determinations. 
The total errors due to the uncertainty of the parameters and the measured spectra range from 0.12 dex for the Ru abundance  to 0.16 for the Mo abundance in hotter stars.
As can be seen in Figs. \ref{mo_teff_} and \ref{ru_teff_}, we found no correlation between the Mo and Ru abundances and \Teff.

\begin{table*}
\caption{
Abundance errors due to the atmospheric parameter uncertainties, for two stars with different set of stellar parameters (\Teff, \logg, and \Vt) - namely, HD154345 (5503, 4.30, 1.3) and HD82106 (4827, 4.10,1.1).
}
\label{errors}
\begin{tabular}{llcccccccc}
\hline
& & HD154345  && & &HD82106 && & \\
 AN & El  & $\Delta$ \Teff+  & $\Delta$ \logg+ & $\Delta$ \Vt+ & tot+ &$\Delta$ \Teff+  & $\Delta$ \logg+ & $\Delta$ \Vt+ & tot+\\
    &     &    [K]           &                 &  [km s$^{-1}$]&      &    [K]          &                & [km s$^{-1}$] &      \\    
\hline
42	&MoI	&0.15 &0.05  &0.01	&0.16	&0.12	&0.05  &0.02	&0.13	 \\ 
44	&RuI	&0.10 &0.03  &0.05	&0.12	&0.11	&0.05  &0.02	&0.12	 \\ 
\hline                             
\end{tabular}
\end{table*}

Unfortunately, no measurements of the Mo and Ru abundances in common stars have been reported elsewhere. 
In Table \ref{comp_han}, we compare our atmospheric parameters and abundances of Mo and Ru with those obtained by  \cite{hansen:14} for two stars common to our sample (\citealt{mishenina:17}).
Overall, the atmospheric parameters derived in both studies are consistent. For HD 22879 ([Fe/H] $\approx$ --1), our upper limit for [Ru/Fe] is consistent with the actual value reported in \cite{hansen:14}.

\begin{table*}
\caption{Comparison of the atmospheric parameters (Mishenina et al. 2017) and Mo and Ru abundances derived in this study with the values reported in \protect\cite{hansen:14} for two common stars.
}
\label{comp_han}
\begin{tabular}{llccccccc}
\hline
HD &   \Teff&  \logg   &  \Vt& [Fe/H]& [Mo/Fe]& [Mo/Fe]& [Ru/Fe]&  reference \\
   &    [K]  &         &[km s$^{-1}$] &     & (3864 \AA)& (5506 \AA)&        &             \\ 
\hline
19445&  5982& 4.38&     1.4& --2.13&     --&          &    0.7  & \cite{hansen:14} \\        
     &  5830& 4.00&     1.1& --2.16&    -- &   --     &         &     ours         \\  
22879&  5792& 4.29&     1.2& --0.95&    -- &   --   &    0.43 &\cite{hansen:14} \\           
     &  5825& 4.42&     0.5&  --0.91&   -- &      0.45&     $>$0.51&        ours  \\									
\hline                             
\end{tabular}
\end{table*}

\section{Results and comparison with theoretical GCE Models}
\label{sec: result, gce}

\cite{hansen:14} compared the behaviour of Mo and Ru with the that of other elements, such as Sr, Zr, Pd, Ag, Ba and Eu, to detect the main sources of these elements in metal-poor stars ([Fe/H] $<$ --0.7). 
They concluded that for the investigated range of [Fe/H], the Mo content is contributed by both the main and weak $s$-processes, the p-process and to a lesser extent by the main and weak $r$-processes. On the other hand, the Ru production is show to be correlated with Ag, suggesting the weak $r$-process to be the main stellar source. 
In this paper, the abundance measurements in F-, G-, and K-dwarfs 
are  representative of the population of stars with higher metallicities compared to those in the sample of \cite{hansen:14}. 

\begin{figure*}
\begin{tabular}{c}
\includegraphics[width=8cm]{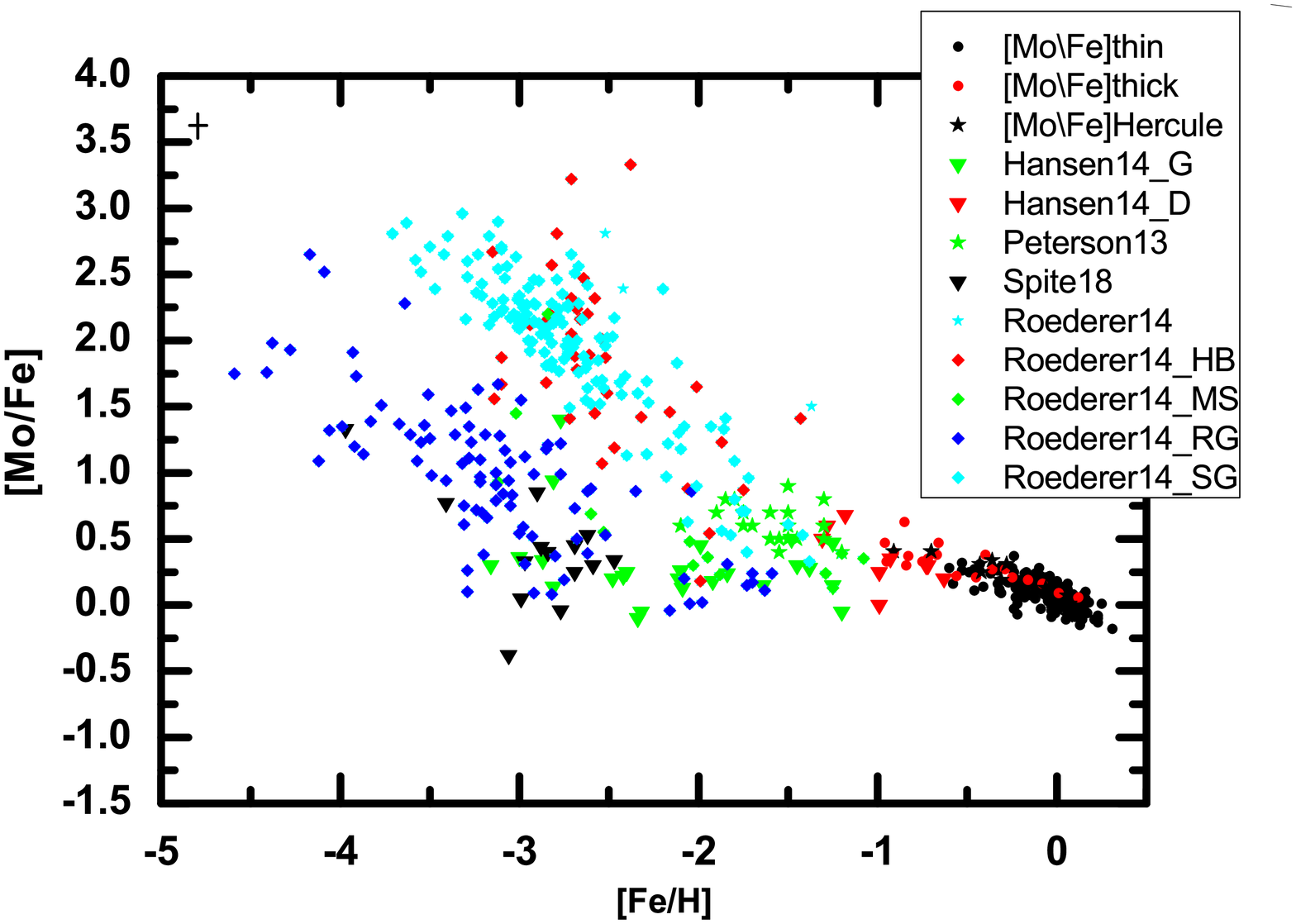}\\
\includegraphics[width=8cm]{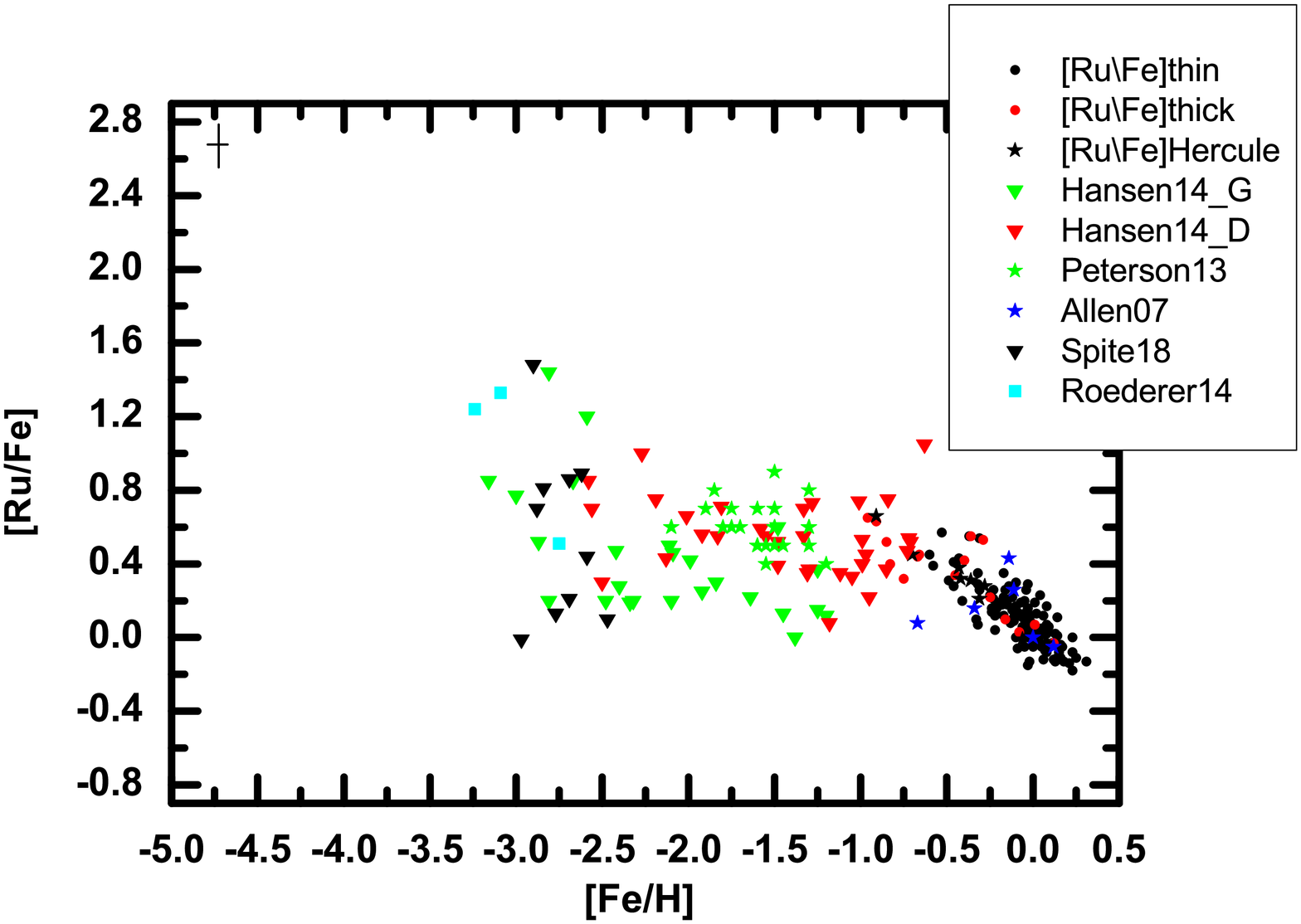}\\
\end{tabular}
\caption{Observed [Mo/Fe] (top panel) and [Ru/Fe] (bottom panel) as a function of [Fe/H] resulted from the comparison of our sample of stellar data with those reported by other authors. Symbols are specified in the figure: Hansen14$\_$G and Hansen14$\_$D refer to giant and dwarf stars by \protect\cite{hansen:14}; Peterson13 are stars by \protect\cite{peterson:13}; Spite18 by \protect\cite{spite:18}, Roderer14$\_$HB, Roederer14$\_$MS, Roderer14$\_$RG and Roderer14$\_$SG are respectively Horizontal Branch, Main Sequence, Red Giant and Subgiant branch stars by \protect\cite{roederer:14a}; Allen07 refers to \protect\cite{allen:07}. The average of the observational errors is provided in the upper left corner.}
\label{morufe_trend1}
\end{figure*}

Fig. \ref{morufe_trend1} shows the [Mo/Fe] and [Ru/Fe] abundance distribution at different [Fe/H], including our determinations for Galactic disc stars and those reported by other authors at different metallicities \citep[][]{allen:07,peterson:13,hansen:14,roederer:14a,spite:18}.
The Mo observations are available for a larger sample of stars at low metallicity as compared to Ru. 
The observational errors for Mo and Ru as follows: 0.1 dex and 0.15 dex, respectively \citep{peterson:13}, 0.15 dex for both Mo and Ru (this work and \cite{hansen:14}), and 0.2 dex for both Mo and Ru \citep[][]{spite:18}. 
\cite{roederer:14a} reported data for 313 stars collected with various telescopes and spectrographs. The authors carried out a thorough analysis and processing of the adopted data, in particular, the parameter estimation, comparison of the equivalent widths obtained with different spectrographs and by different researchers, as well as application of the atmospheric models, calculation codes, line lists, etc. The comparison of the Mo and Ru abundances estimated by \cite{roederer:14a} with those obtained by other authors for all the sample stars has shown uncertainties ranging from 0.2 -– 0.3 dex to 0.4 dex for Mo and Ru, respectively. Note that there are no NLTE calculations for Mo or Ru currently available. 
With regard to our sample of stars, since we use weak subordinate lines, and they are formed in the deep atmospheric layers wherein collisions with electrons create (establish) the LTE conditions, the NLTE corrections should be negligible and leveled using our analysis relative to the Sun. They should not make a significant contribution to the errors in the measurements reported in this paper. 
For more metal-poor stars, they could yield more relevant corrections. 
The correlations between Mo, Ru, Y, Zr, Ba, Sm, Eu \citep{mishenina:13}, and Sr \citep{mishenina:19} are illustrated in Figs. \ref{elmo_n}, and Figs. A1-A3 in the Appendix. The slopes and errors obtained for our stellar sample are summarized in Table~\ref{slopes}. We cannot deduce any detailed information from these slopes without GCE simulations representative of the disc stars. However, using the data shown in the figures, we can draw several important conclusions.  
The Mo and Ru abundance trends with respect to the $r$-process element Eu (Figs.\ref{elmo_n}, \ref{elru}) 
show no clear correlations for the thin disc stars. Moreover, the Mo enrichment does not correlate closely with Ru. 
Such a pattern could be associated with a late nucleosynthesis source yielding Ru, but not producing efficiently other elements in the same mass region as Mo. It is the most likely that such an extra source would not be an $s$-process source, since Mo and Ru have similar patterns of the $s$-process production (see Table \ref{diff_sun}). Having analyzed the correlation between  various elements at the near solar metallicity we can derive that the galactic stellar sources which contribute to Mo and Ru content are at least partially different. Furthermore, we confirm that the contribution of the main $s$-process to the Mo and Ru solar abundances is lower than that for the $s$-process elements such, as Sr, Y and Ba. An in-depth study of the Mo and Ru production, as well as the relative impact of different stellar sources, with application of detailed GCE simulations is required. Note that the observational uncertainties are similar for Mo and Ru as discussed in the previous sections. Therefore, they cannot explain such different behaviour observed for Mo and Ru.

\begin{figure*}
\begin{tabular}{c}
\includegraphics[width=11cm]{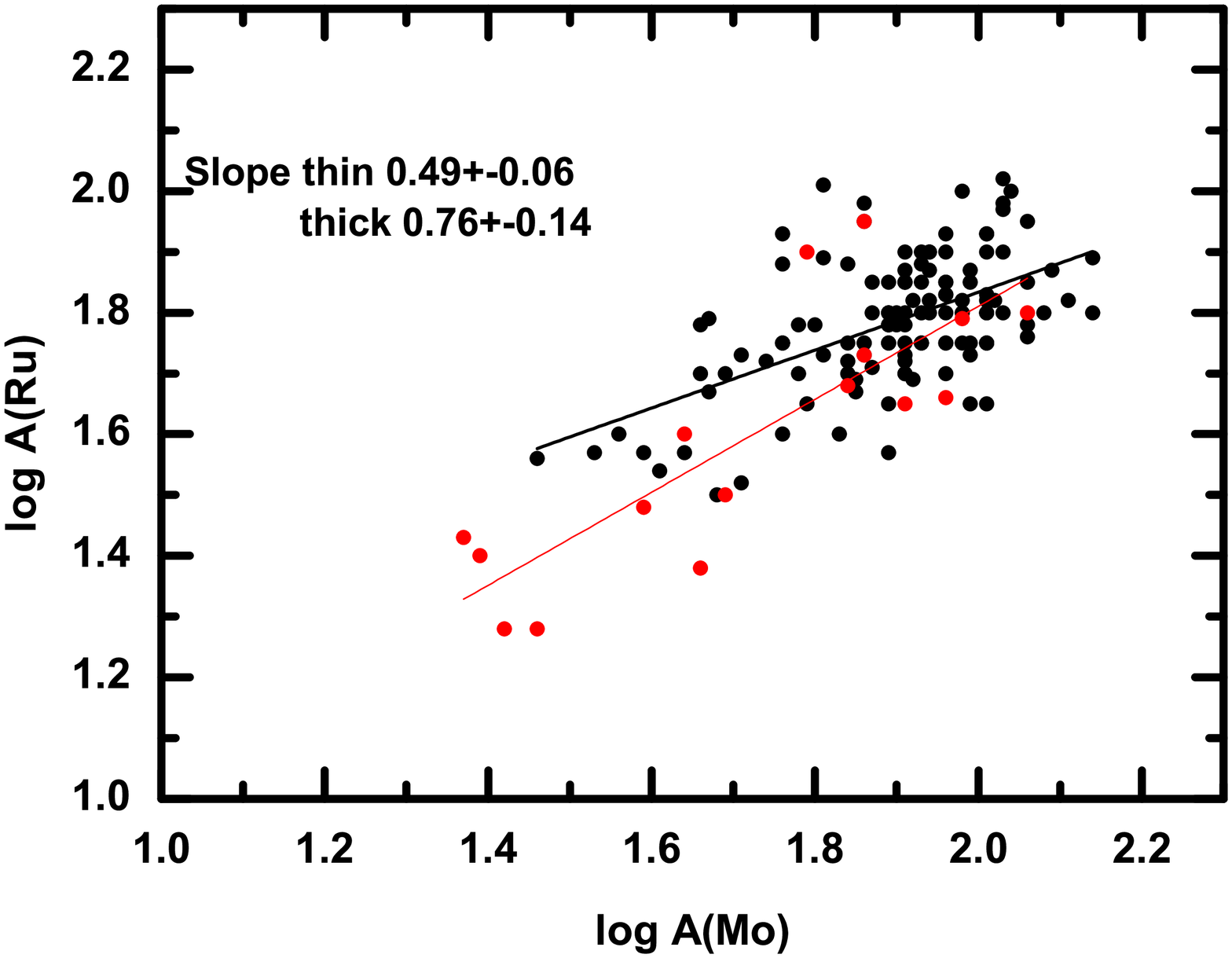}\\
\includegraphics[width=11cm]{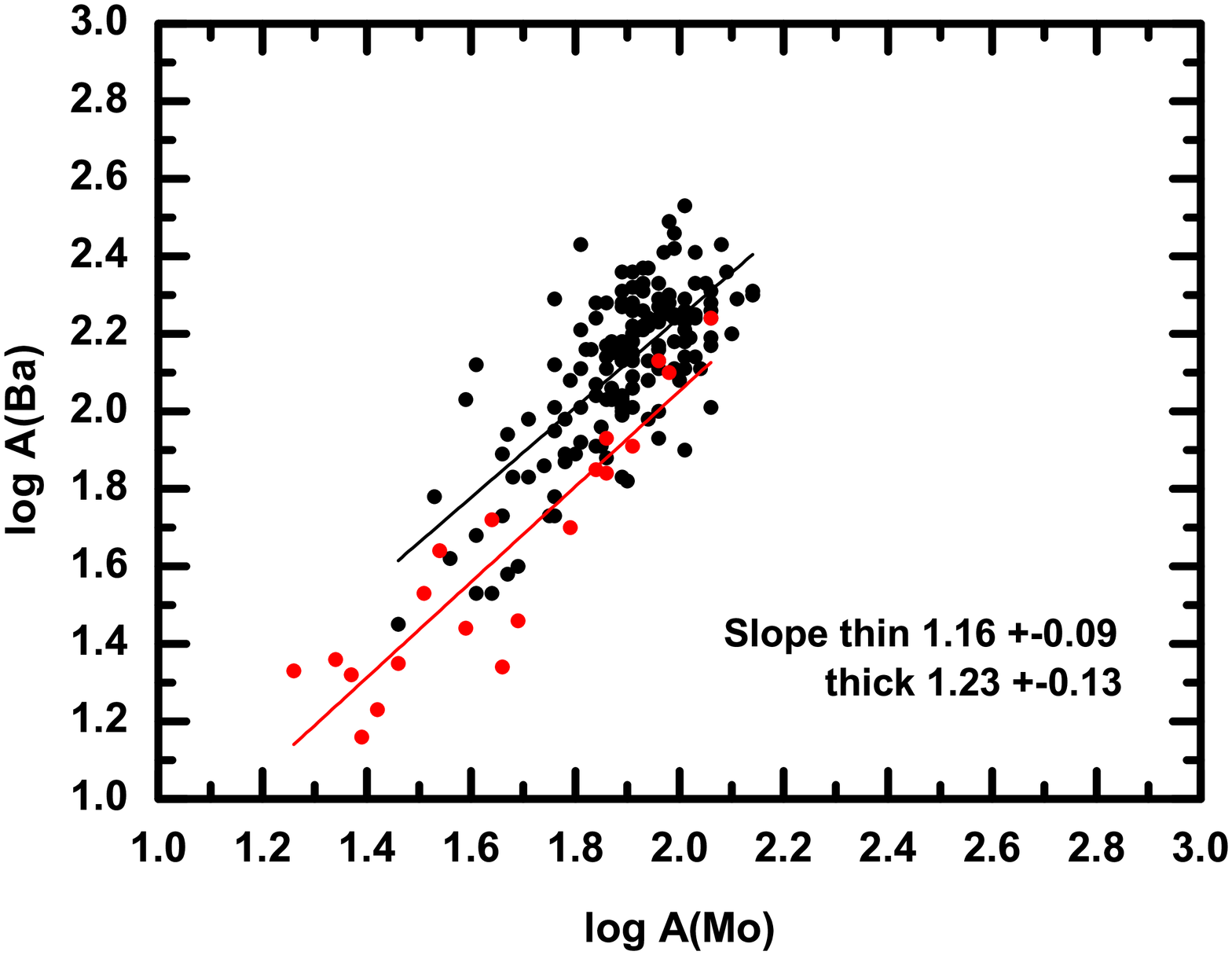}\\
\includegraphics[width=11cm]{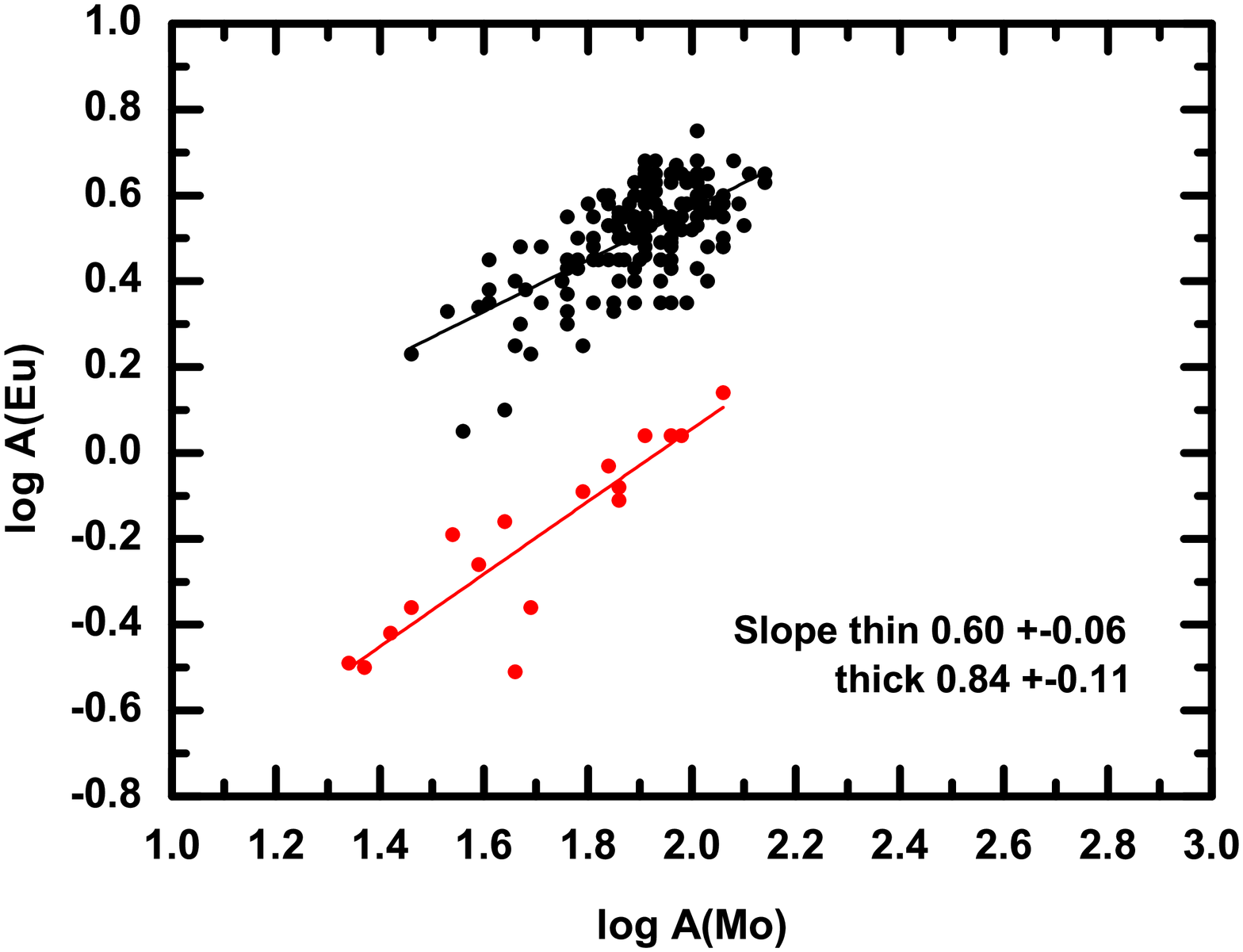}\\
\end{tabular}
\caption{Trends of log A(El) where El = Ru, Ba, and Eu vs. log A(Mo) for thin disc stars (small black circles) and thick disc stars (red circles). }
\label{elmo_n}
\end{figure*}

\begin{table*}
\caption{Elemental abundance trends related to Mo and Ru, as shown in Figs. \ref{elmo_n}, A1, A2, A3, for thin disc stars (3rd column) and thick disc stars (4th column).}
\label{slopes}
\begin{tabular}{cccc}
\hline
\hline
   1  & 2  &    3 &            4\\        
\hline
Element & Reference  & Slope $\pm$ Error & Slope $\pm$ Error \\
\hline
Ru	& Mo	& $0.49\pm0.06$  & $0.76\pm0.14$ \\ 
Ba	& Mo	& $1.16\pm0.09$  & $1.23\pm0.13$ \\ 
Ba	& Ru	& $0.99\pm0.15$  & $1.32\pm0.26$ \\
Sr	& Y 	& $1.17\pm0.04$  & $1.07\pm0.04$ \\ 
Sr	& Mo 	& $1.48\pm0.08$  & $1.36\pm0.12$ \\ 
Sr	& Ru 	& $1.15\pm0.16$  & $1.30\pm0.30$ \\ 
Y	& Mo	& $0.97\pm0.08$  & $1.27\pm0.11$ \\
Zr & Mo & $0.83\pm0.08$  & $0.97\pm0.08$ \\ 
Sm	& Mo 	& $0.67\pm0.06$  & $1.06\pm0.10$ \\ 
Sm	& Ru 	& $0.64\pm0.09$  & $1.17\pm0.19$ \\ 
Eu	& Mo 	& $0.60\pm0.06$  & $0.84\pm0.11$ \\ 
Eu	& Ru 	& $0.43\pm0.10$  & $0.80\pm0.17$ \\ 
\hline  
\hline                           
\end{tabular}
\end{table*}

The application of GCE models allows us to take into account the contribution of various nucleosynthesis sources occurring at different timescales during the evolution of the elements. 
GCE simulations serve as a fundamental tool to understand the complex history of enrichment of elements like Mo and Ru. Recently, \cite{prantzos:18} have carried out the investigation of the chemical evolution of the elements from H to U in the Milky Way halo and local disc. The authors used metallicity-dependent yields from low- and intermediate-mass stars (LIM, AGB), and from rotating massive stars.  
They found that the solar isotopic composition of pure $s$-process isotopes could be reproduced within 10\,\% accuracy.
They also reproduced the $s$-process abundances of isotopes for 90 $<$ A $<$ 130 \citep[solar LEPP][]{montes:07}. 
The differences between their findings and those resulted from the GCE simulations reported by  \cite{bisterzo:17} were mainly due to the yields adopted for rotating massive stars, in particular, \cite{prantzos:18} used the isotopic yields from \cite{limongi:18}. Moreover, the AGB yields used in both studies were not similar that could affect the results obtained for the elements which are subject to our study. \cite{prantzos:18} have also concluded that there are significant differences at lower [Fe/H] for which the chemical evolution is mainly governed by massive stars and emphasized some deficiency in Zr and Mo.

\begin{figure*}
\begin{tabular}{c}
\includegraphics[width=14cm]{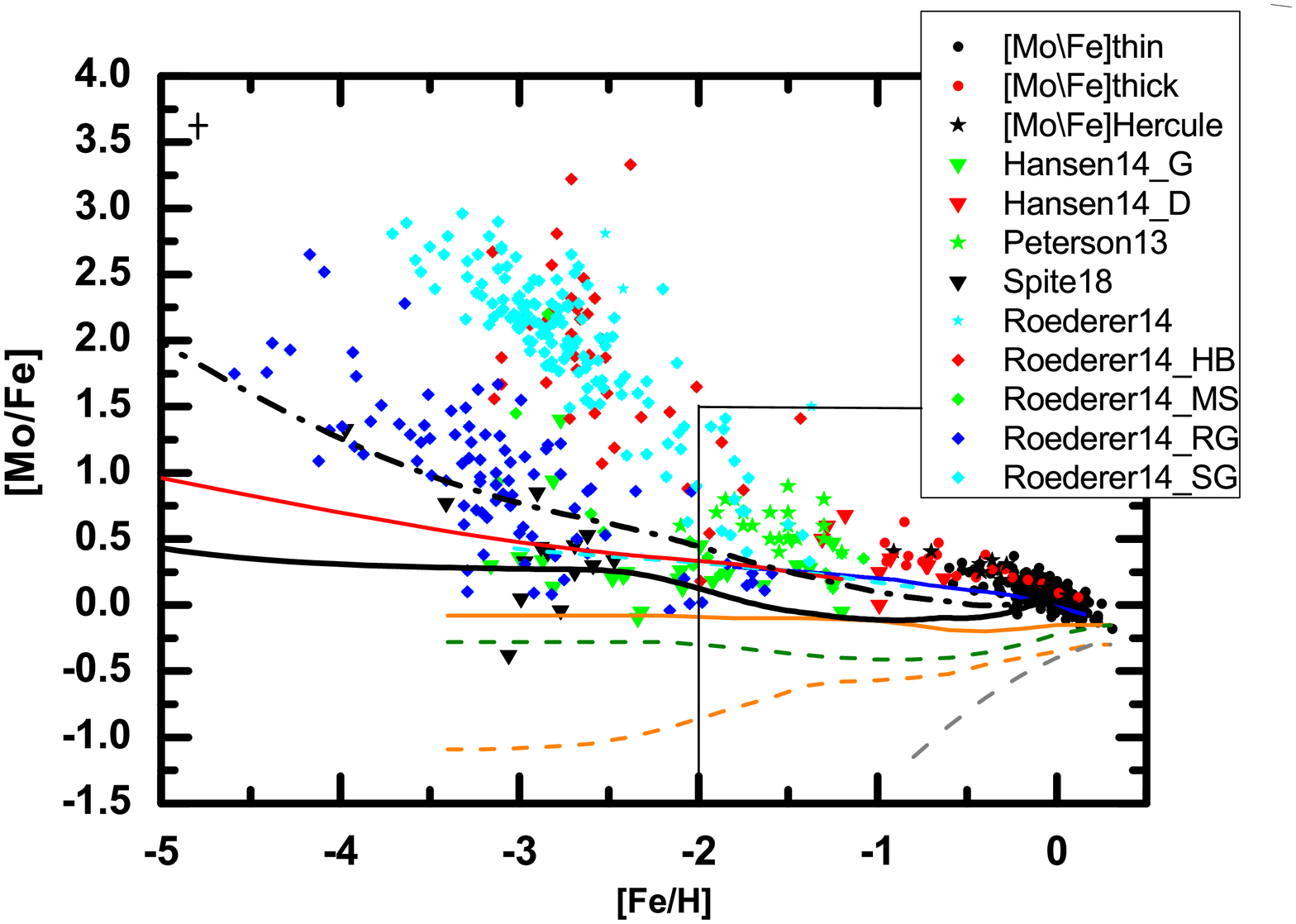}\\
\includegraphics[width=14cm]{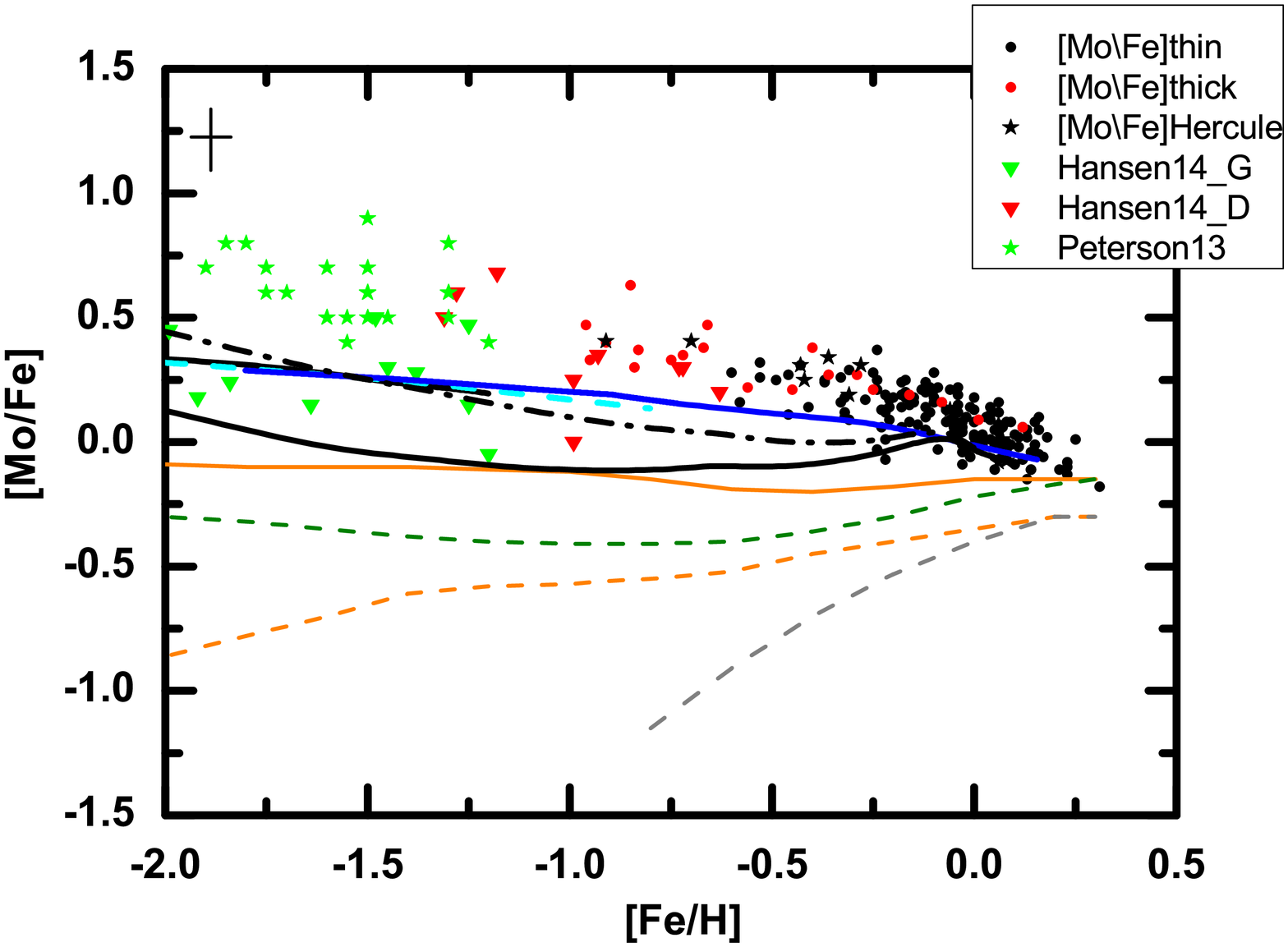}\\
\end{tabular}
\caption{Top Panel: evolution of [Mo/Fe] as a function of [Fe/H] predicted by the chemical evolution models of \protect\cite{prantzos:18} and \protect\cite{travaglio:04} as compared to the observation data (markers are specified in the figure).
For \protect\cite{prantzos:18} simulations, the following models are shown:
i) LIM stars, rotating massive stars plus their fiduciary $r$-process (the baseline model, orange solid curve); ii) LIM stars, non-rotating massive stars and $r$-process (green dashed curve); iii) LIM stars and non-rotating massive stars without $r$-process contribution (gray dashed curve); and iv) LIM stars plus rotating massive stars without the $r$-process contribution (orange dashed curve). 
For the \protect\cite{travaglio:04} predictions, the models for halo, thick and thin disc of the Milky Way are reported (solid red, dashed cyan and solid blue curves, respectively). The curves overlap within the metallicity range --2 $<$ [Fe/H] $<$ --1. The prediction by the OMEGA+ code,  short delay time ($r$ - process) and delay time distribution ($r$ - process) marked as black dots - dashed line and solid line, respectively. Bottom panel: the same as on the top panel, but for the metallicity range of the Galactic disc. The average observational errors is provided in the upper-left corner.} 
\label{mofe_trend}
\end{figure*}

\begin{figure*}
\begin{tabular}{c}
\includegraphics[width=14cm]{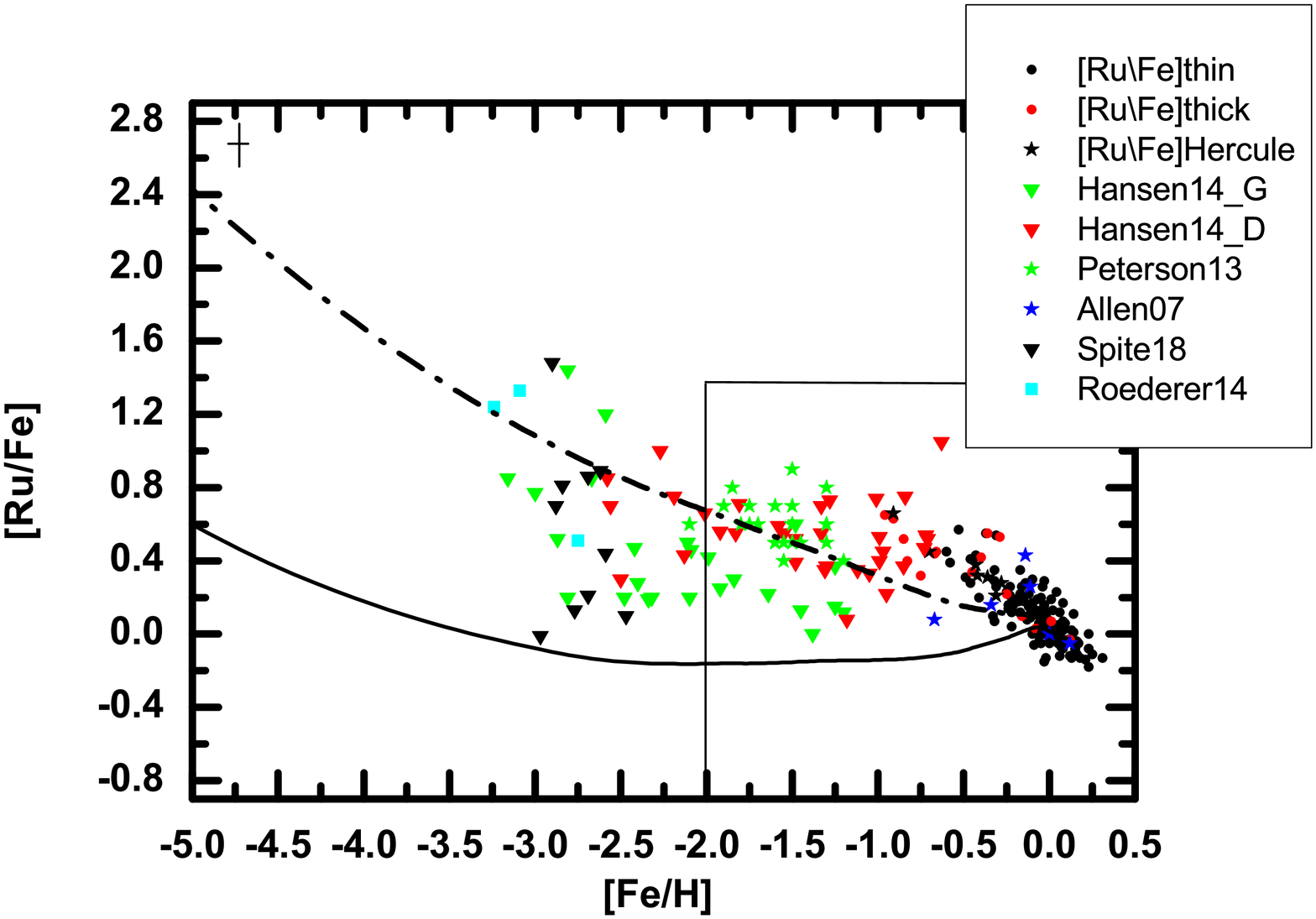}\\
\includegraphics[width=14cm]{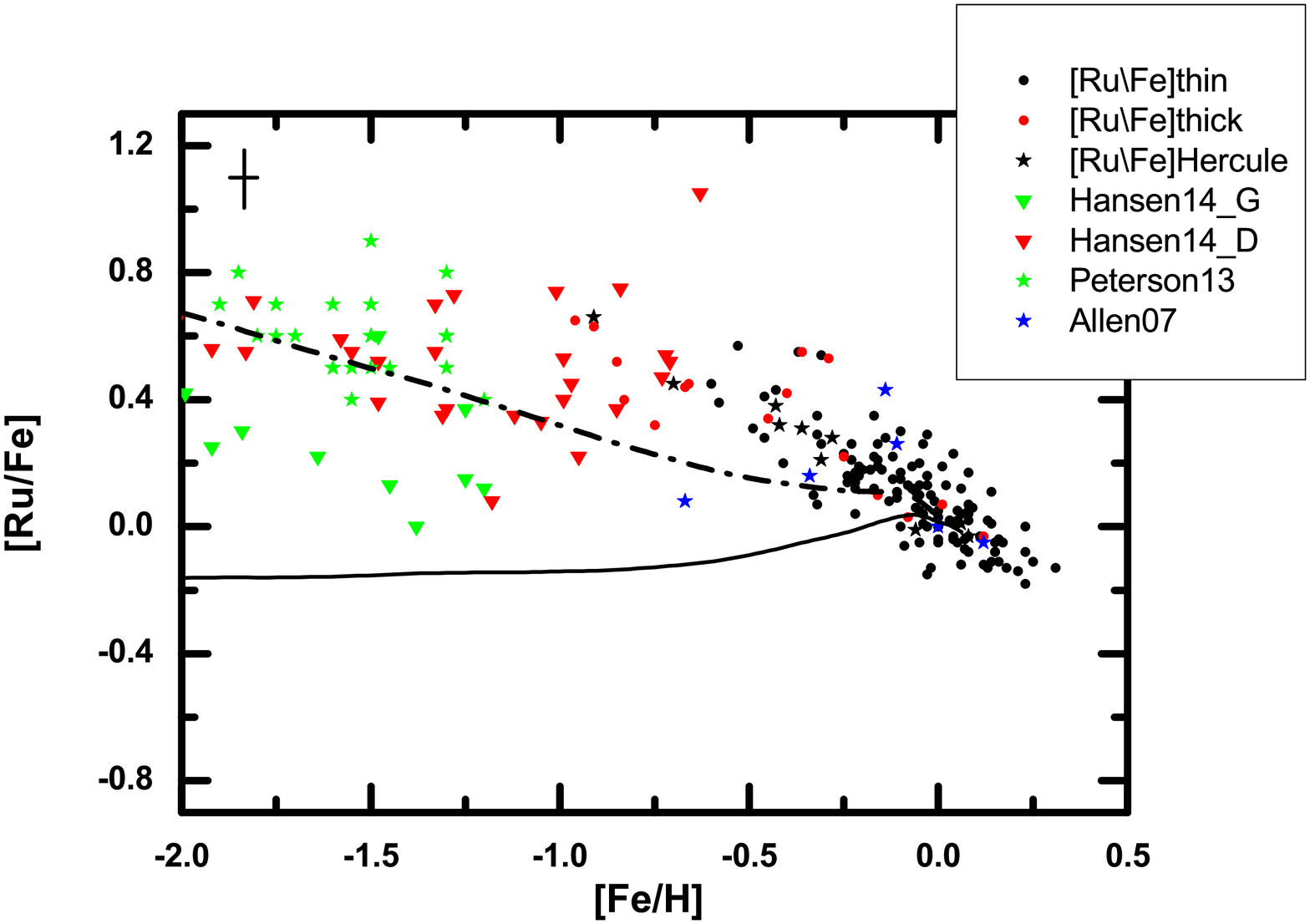}\\
\end{tabular}
\caption{Top panel: evolution of [Ru/Fe] as a function of [Fe/H] predicted by  by the OMEGA+ code, the short delay time ($r$ - process) and delay time distribution ($r$ - process) marked with black dot, dashed and solid line, respectively, as compared to the observational data (markers are specified in the figure). Bottom panel: the same as on the top panel, but for the metallicity range of the Galactic disc. The average observational errors is provided in the upper-left corner.}
\label{rufe_trend}
\end{figure*}

Our new observational data for Mo and Ru along with those reported by \cite{hansen:14, peterson:13,roederer:14a,spite:18} are presented in Figs. \ref{mofe_trend} and \ref{rufe_trend}, respectively. The bottom panels in the figures illustrate the evolution of Mo and Ru for the metallicity range of the Galactic disc.
The GCE evolution of Mo predicted by \cite{prantzos:18} (as presented in Fig.16 in their study) is compared with the observational data in Fig. \ref{mofe_trend}.
As also highlighted by the authors, the theoretical trends are not reproducing the Galactic behavior of Mo: GCE simulations do not produce enough Mo compared to Fe in comparison to observations. 
Fig. \ref{mofe_trend} also shows the GCE results of \cite{bisterzo:17}, who used the same chemical evolution model as \cite{travaglio:04}. The latter prediction for [Mo/Fe] assumed 40\% $s$-process contribution from AGB stars \cite[see][for details]{bisterzo:17}, 10\% contribution from the $r$-process, and 1\% contribution from the weak $s$-component from massive stars. We also obtained 49\% contribution from LEPP (derived from \cite{travaglio:04}). As in \cite{travaglio:15}, we can also derive separately the $p$-process contribution to the two $p$-only isotopes of Mo, i.e. $^{92, 94}\mathrm{Mo}$ from Type Ia supernovae (a single degenerate scenario). However, the $p$-process contribution to the total Mo abundance is irrelevant for reproducing the Mo observations in the Galaxy.

In Figs. ~\ref{mofe_trend} and ~\ref{rufe_trend} we also show the evolution of [Mo/Fe] and [Ru/Fe], as predicted using the open-source GCE code \texttt{OMEGA+} (\citealt{cote:18c}), which is part of the JINAPyCEE Python package\footnote{https://github.com/becot85/JINAPyCEE}.
This is a two-zone model consisting of a classical one-zone chemical evolution model located at the center of a large gas reservoir (the circumgalactic medium of the simulated galaxy). For low- and intermediate-mass stars, we used the stellar yields reported in \citealt{cristallo:15} with no rotation and standard $^{13}\mathrm{C}$  pocket, which are available with the
 the F.R.U.I.T.Y\footnote{http://fruity.oa-teramo.inaf.it/modelli.pl} database. 
For thermonuclear supernovae (SNIa), we adopted the yields from \cite{iwanomoto:99} and distributed them in time following a function based on the observed delay-time distribution function for SNIa (see \citealt{2016ApJ...824...82C} and \citealt{2018ApJS..237...42R} for more details).
For the CCSNe yields, we used the NuGrid massive star models (\citealt{2018MNRAS.480..538R}) along with the \textit{delayed} supernova engine prescription (\citealt{2012ApJ...749...91F}). In order to calculate the integrated stellar yields used in the GCE simulations, we did not use the 12\,M$_\odot$ models at all metallicities. The SN explosion setup used for these models causes an overproduction of Fe abundances when compared to the solar composition, which indicates that the conditions obtained are not representative of those in most of 12\,M$_\odot$ CCSNe do (see, \citealt{cote:18a,2018ApJ...861...40P}).
We also did not consider the 15\,M$_\odot$ model at $Z=0.006$. That single model included a strong $\alpha$-rich freezout contribution \citep[e.g.][]{woosley:92,pignatari:16b}, that resulted in the overestimated  GCE production for some first-peak neutron-capture elements, such as Y and Zr, in our simulations.
Therefore, the $\alpha$-rich freezout component obtained in that model is not representative of what 15\,M$_\odot$ CCSNe stars typically produce.
For the GCE models considered below, for simplicity, we replaced the 15\,M$_\odot$ model at $Z=0.006$ with the 15\,M$_\odot$ model at $Z=0.001$, without causing any impact on the GCE of Mo and Ru. The only difference between the two \texttt{OMEGA+} models presented in Figs. ~\ref{mofe_trend} and ~\ref{rufe_trend}, is a different setup for the $r$-process production. For both models, the dominant $r$-process source are neutron star mergers \citep[e.g.][]{cowan:19,cote:19}. However, for each $r$-process event we assume that the ejecta is released either 30\,Myr after the formation of the progenitor star (short-delay time setup), or released following a delay-time distribution function in the form of $t^{-1}$ from 10\,Myr to 10\,Gyr (delay-time distribution setup) \citep{chruslinska:18}.

As can be seen in the bottom panel of Fig. \ref{mofe_trend}, the contribution from massive rotating stars or from AGB stars at higher metallicities does not solve the issue of underproduction in theoretical predictions as compared to the observations \citep[][]{prantzos:18}. 
The simulations by \cite{travaglio:04} and \texttt{OMEGA+} (short delay time setup) seem to show a better consistency with Mo observations, reducing the average underproduction to about 0.1 dex. For the data reported in \cite{travaglio:04} this might be due to the additional contribution to Mo by the Lighter Element Primary Process component, considered in these GCE models. Concerning \texttt{OMEGA+} results, the difference is mainly due to the $r$-process sources different from those in two other sets of GCE simulations. In particular, the yields of neutron-star mergers are implemented as decoupled with CCSNe which is the main source of Fe at low metallicity.

According to the results of our comparison of  the \texttt{OMEGA+} and  \cite{travaglio:04} more thoroughly, the $s$-process contribution to the solar abundances of Mo and Ru is 60\% and 45\%, respectively, which is different from 40\% and 24\% obtained by \cite{travaglio:04}.
The $r$-process contribution obtained is 16\% and 45\% respectively, compared to 12\% and 50\% derived from the elemental distribution of the $r$-process in the metal-poor star CS 22892-052 and used as a reference of the $r$-process contribution in \cite{travaglio:04}.
Despite such a higher $s$-process contribution obtained in \texttt{OMEGA+} calculations, the requirement for having additional sources for Mo at low metallicity is consistent with the results of other GCE simulations referred to in this paper. 
Concerning Ru, the two \texttt{OMEGA+} models show different results with a higher [Ru/Fe] trend using the short delay time setup. On the other hand, even in the most optimistic conditions, at metallicities lower than solar ones, the GCE model yield is 0.2 dex lower as compared to the observations.
 
\begin{figure*}
\begin{tabular}{c}
\includegraphics[width=16cm]{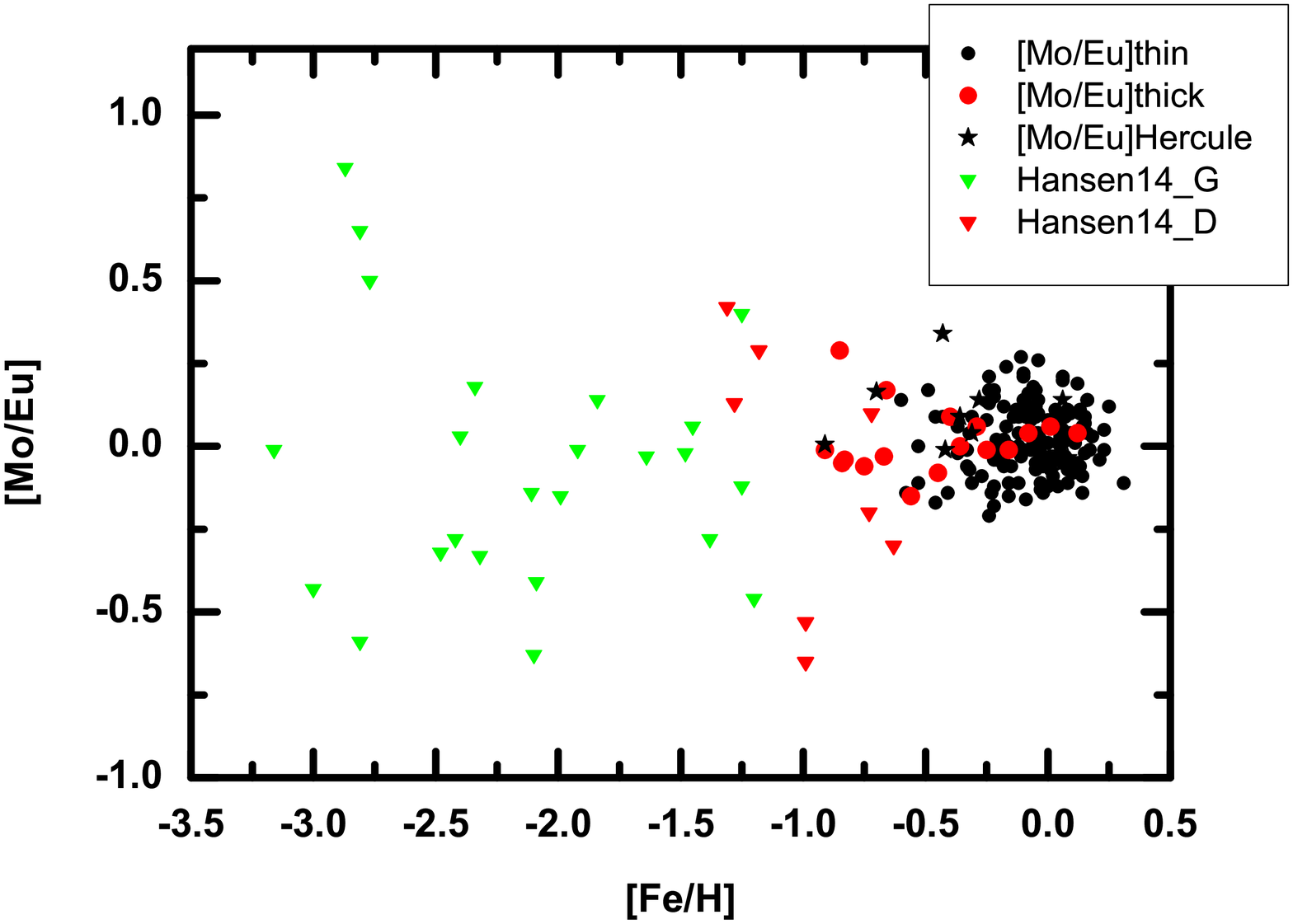}\\
\end{tabular}
\caption{Evolution of [Mo/Eu] as a function of [Fe/H].}
\label{mo_eu}
\end{figure*}

Despite the fact that the $r$-process contributes to a small fraction of the solar Mo, and a half of that of Ru, it becomes more significant at low metallicities, where the $s$-process contribution from AGB stars becomes marginal. In particular, we show in the bottom panels of Figs. \ref{mofe_trend} and \ref{rufe_trend} that the properties of the $r$-process sources adopted in the GCE simulations have a strong impact on the abundances of Mo and Ru for [Fe/H] $<$ -0.2 dex,  if we assume that all $r$-process events carry a solar $r$-process residual pattern for the yields \citep[e.g.][]{arnould:07}. In a more general sense, the study of the chemical evolution of these elements can provide additional new constraints on the $r$-process production in the Galaxy and other nucleosynthesis processes active in the early Galaxy. 
In Fig. \ref{mo_eu}, we show that while [Mo/Eu] is consistent within 0.4 dex in our stellar sample, the observed scatter increases up to about 2 dex in more metal poor stars \citep[][]{hansen:14}. This would imply that at least Mo-poor and Mo-rich nucleosynthesis sources with respect to Eu were active in the early Galaxy. At the same time, the [Mo/Fe] and [Ru/Fe] scatter observed at low metallicities (see Fig. \ref{morufe_trend1}) is quite similar to that of [Eu/Fe], indicating that also the production of Mo and Ru with respect to Fe at low metallicity is associated to rare events. A detailed study of these observations for [Fe/H] $\lesssim$ -2 would possibly require an inhomogeneous galactic chemical evolution study \citep[e.g.][]{wehmeyer:15,mishenina:17}.

\section{Conclusions}
\label{sec: conclusions}

We presented new observational data for the light trans-Fe elements Mo (Z = 42) and Ru (Z = 44) in F-, G-, and K-stars belonging to the substructures of the Galaxy with metallicities ranging from -–1.0 $<$ [Fe/H] $<$ +0.3. The spectra of Galactic disc stars have a high resolution of 42,000 and 75,000 and signal-to-noise ratio better than 100. The Mo and Ru abundances were derived by comparing the observed and synthetic spectra in the region of Mo I lines at 5506, 5533 \AA~ (for 209 stars), and Ru I lines at 4080, 4584, and 4757 \AA~ (for 162 stars) in the LTE approximation. For all the stars the Mo and Ru abundance determinations were obtained for the first time. Taking into consideration the observational data reported in other studies at low metallicities, this work enables us to analyse the complete trend of Mo and Ru abundances in the Milky Way.

As follows from the observations at lower metallicities, the existing GCE models with the nucleosynthesis sites and stellar yields included therein underproduce Mo and Ru compared to observational data also in the Galactic disc.
Canonical stellar sources of heavy elements, such as the $s$-process in massive stars and AGB stars or the $r$-process, do not appear to produce sufficient amount of these elements. Factoring in the additional Lighter Element Primary Process or LEPP in GCE simulations allows to obtain a better fit. However, the nature and the origin of LEPP is still a matter for debate, and such a zoo of numerous independent stellar processes could contribute instead to the production of various elements. 
According to the GCE models presented in this paper, also the assumption of an $r$-process source disentangled by CCSNe like neutron star mergers can provide in principle a better fit for the observations. However, even the most Mo-rich and Ru-rich GCE simulations cannot reproduce all the observed Mo and Ru. Similar indications seem to be obtained for metal-poor stars, but hydrodynamics chemical evolution or inhomogeneous chemical evolution models are needed in order to study the inhomogeneous enrichment of the galactic halo.

In summary, the origin of the two elements remains an open question requiring further detailed studying. We found that some other stellar sites and their contributions should be included in the GCE calculations, apart from the classical nucleosynthesis processes. As regards Eu, the large scatter observed for [Mo/Fe] and [Ru/Fe] at low metallicities would be consistent with the contribution from a rare stellar source.
For the thick and thin disc stars in our sample, we found that the Mo enrichment is correlated with both Ba and Eu. On the other hand, Ru shows a much higher dispersion with respect to Mo, Ba and Eu. A possible scenario that we suggested and discussed is that Ru could be efficiently produced by an extra stellar nucleosynthesis source active in the Galactic disc. Further investigation with GCE simulations is required to better define the nature of such a source. Today, we can only argue that it is not an s-process source, since the s-process contributions to Mo and Ru are similar. At present, spectroscopic abundance measurements available for Mo and Ru are based on the LTE calculations with no NLTE corrections currently available. Though it should not be an issue within the metallicity range of our stellar sample, more significant corrections could be required for the observations in metal-poor stars. However, since the discrepancy between theoretical predictions and observations is already evident from the simulations of the chemical evolution of the Milky Way disc, it would not affect the main findings and conclusions presented in this paper

\section*{Acknowledgements}

This paper was  based  on  the observation data  collected at OHP  Observatory, France. TM, TG, MP, FKT grateful for the support from the Swiss National Science Foundation, project SCOPES No. IZ73Z0$_{}$152485.
MP acknowledges significant support to NuGrid from NSF grant PHY-1430152 (JINA Center for the Evolution of the Elements) and STFC (through the University of Hull's Consolidated Grant ST/R000840/1), and access to {\sc viper}, the University of Hull High Performance Computing Facility.
MP acknowledges the support from the "Lendület-2014" Programme of the Hungarian Academy of Sciences (Hungary), and from the BRIDGCE UK network.
FKT acknowledges support from the European Research Council (FP7) under ERC 
Advanced Grant Agreement 321263 FISH. BC and MP acknowledges support from the ERC Consolidator Grant (Hungary) funding scheme (project RADIOSTAR, G.A. n. 724560) and from the National Science Foundation (USA) under grant No. PHY-1430152 (JINA Center for the Evolution of the Elements).
This article is based upon work from the ChETEC COST Action (CA16117), 
supported by COST (European Cooperation in Science and Technology). TM thanks to S. Korotin for discussions. The authors appreciate very useful comments provided by the anonymous referee.

\bibliography{molybdenum}

\appendix
\section{}
\begin{figure*}
\begin{tabular}{c}
\includegraphics[width=11cm]{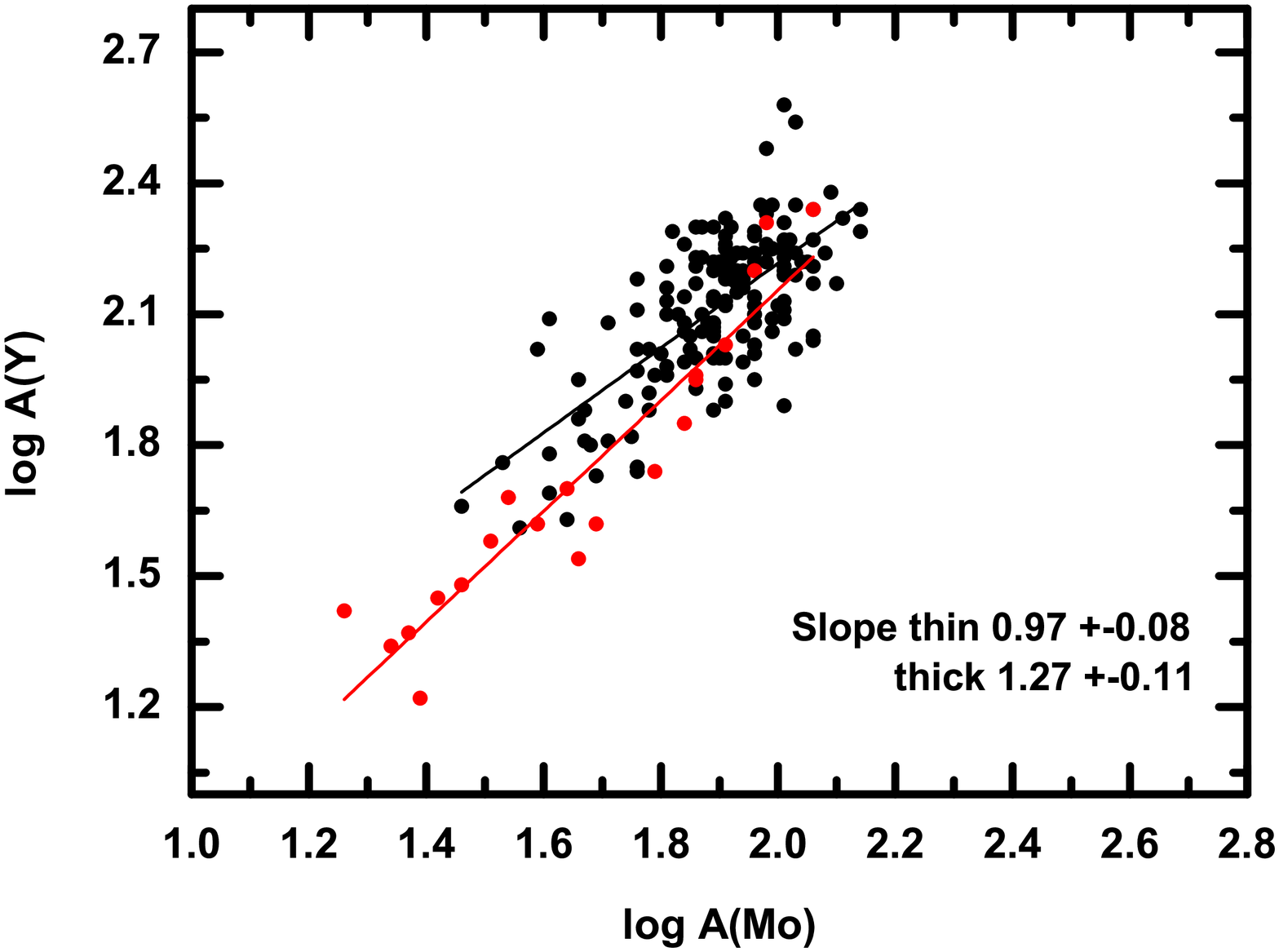}\\
\includegraphics[width=11cm]{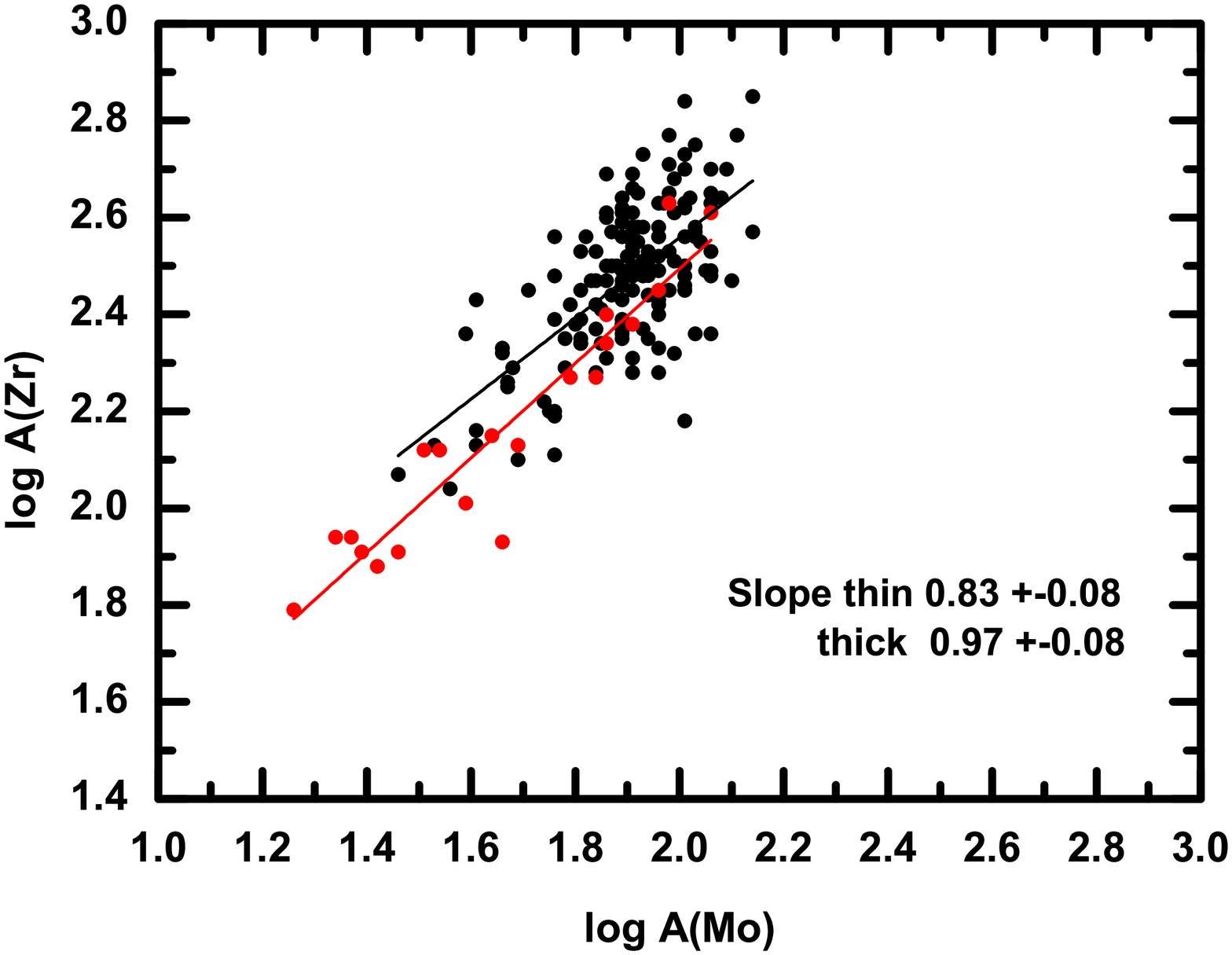}\\
\includegraphics[width=11cm]{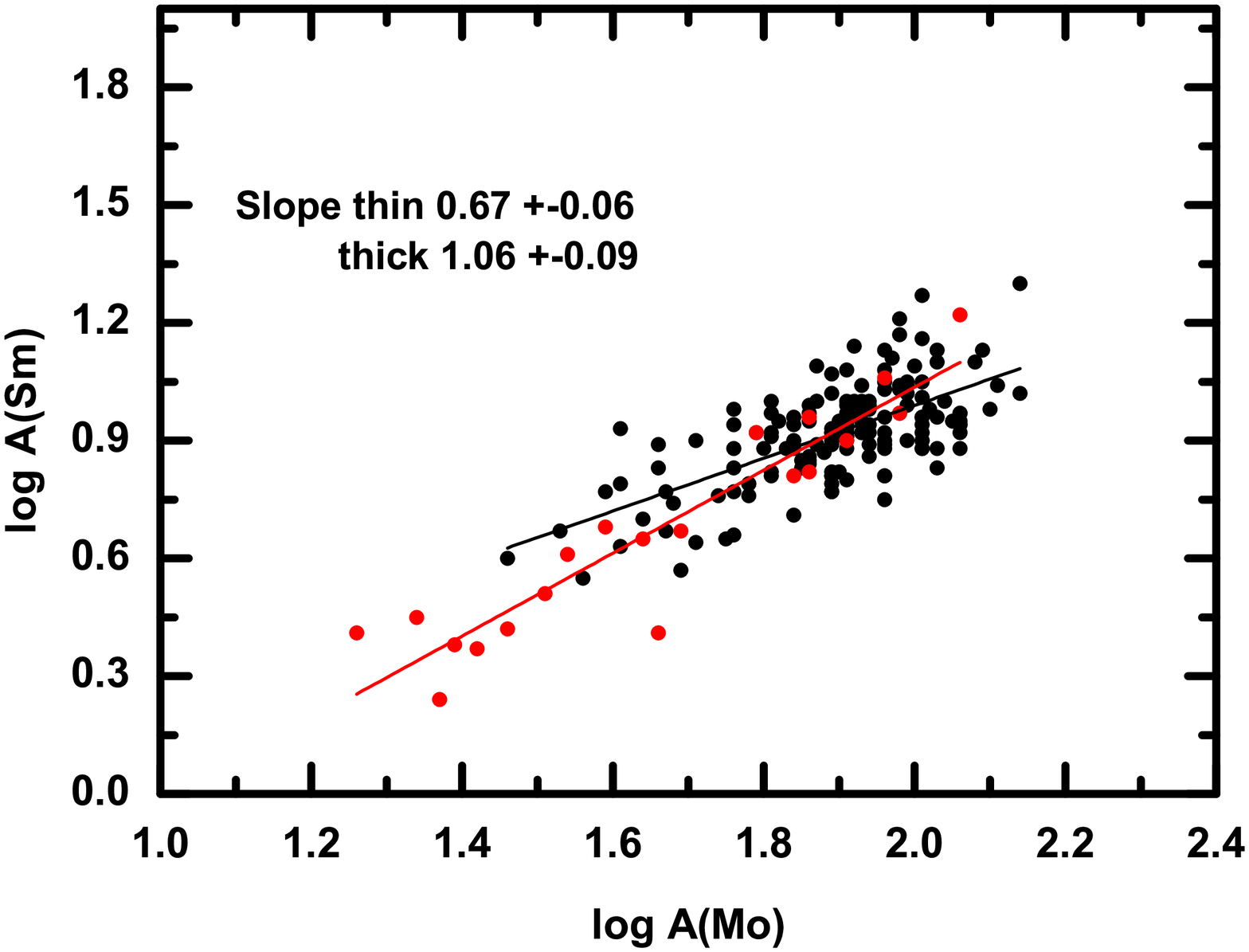}\\
\end{tabular}
\caption{Trends for log A(El) where El = Y, Zr, Sm  with respect to log A(Mo). Notation is the same as presented in Fig. \ref{elmo_n}.}
\label{elmo1_n}
\end{figure*}

\begin{figure*}
\begin{tabular}{c}
\includegraphics[width=11cm]{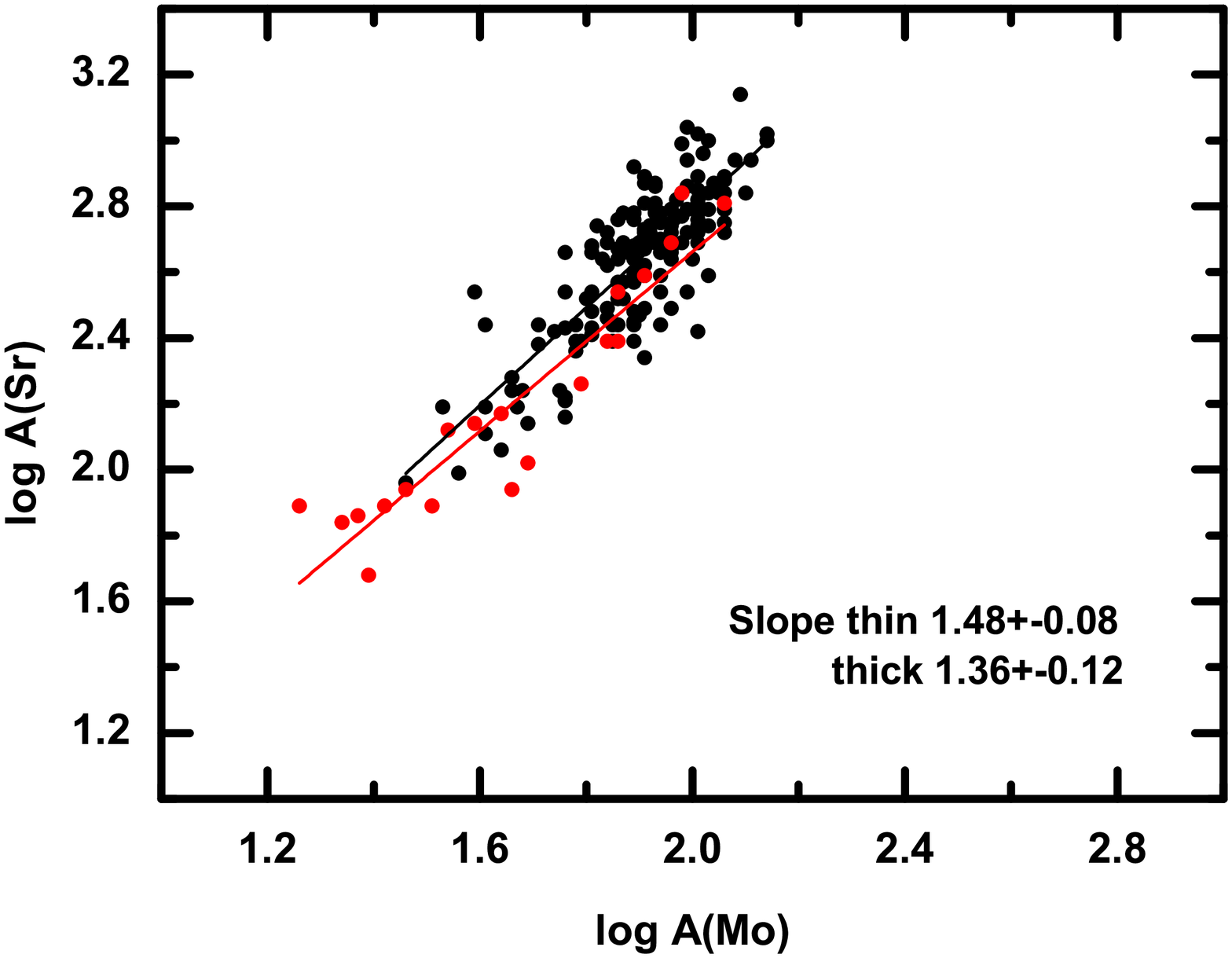}\\
\includegraphics[width=11cm]{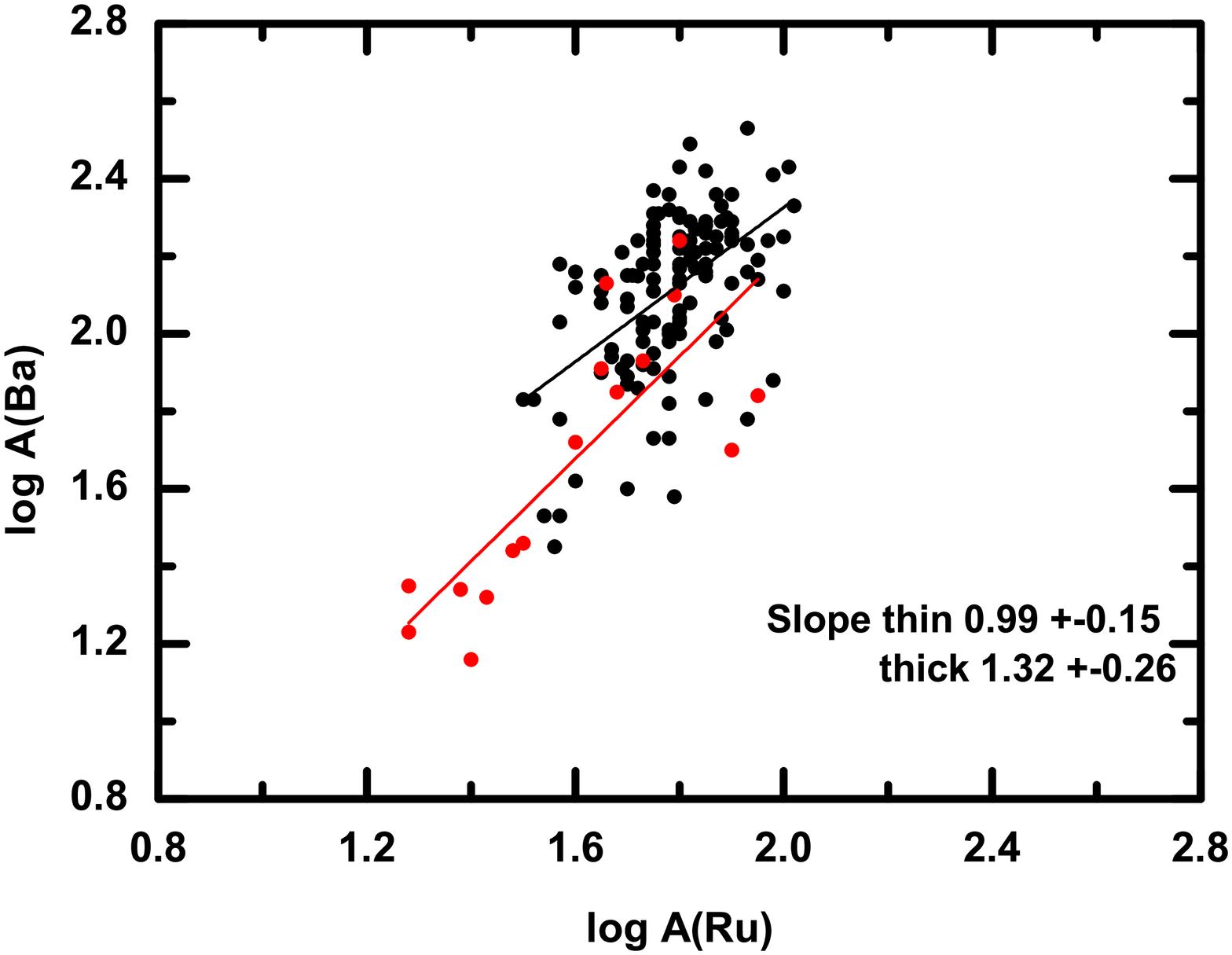}\\
\includegraphics[width=11cm]{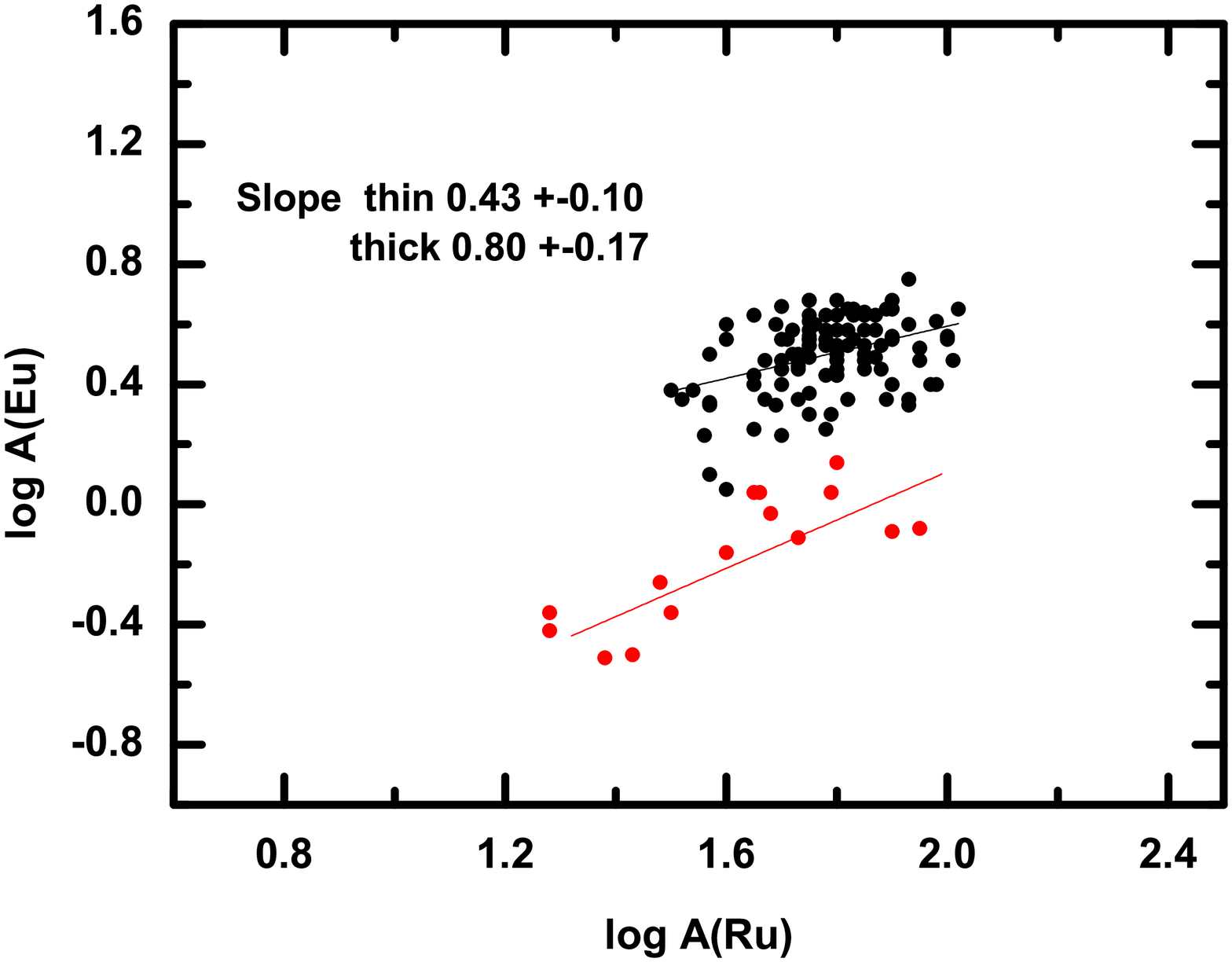}\\
\end{tabular}
\caption{Trends of logA(Sr) vs. logA(Mo) and log A(El) where El = Ba, and Eu vs. log A(Ru). Notation is the same as presented in Fig. \ref{elmo_n}.}
\label{elru}
\end{figure*}

\begin{figure*}
\begin{tabular}{c}
\includegraphics[width=11cm]{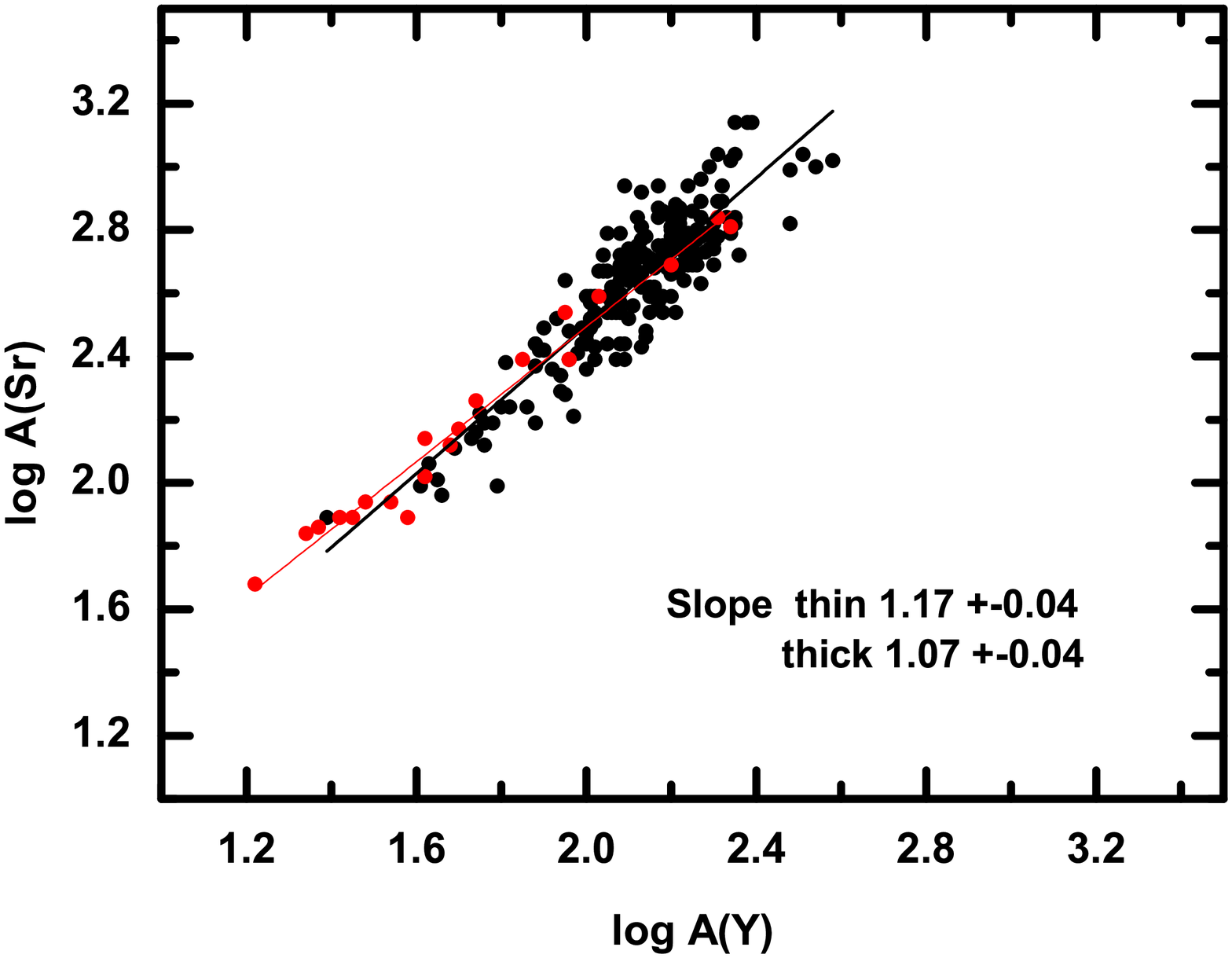}\\
\includegraphics[width=11cm]{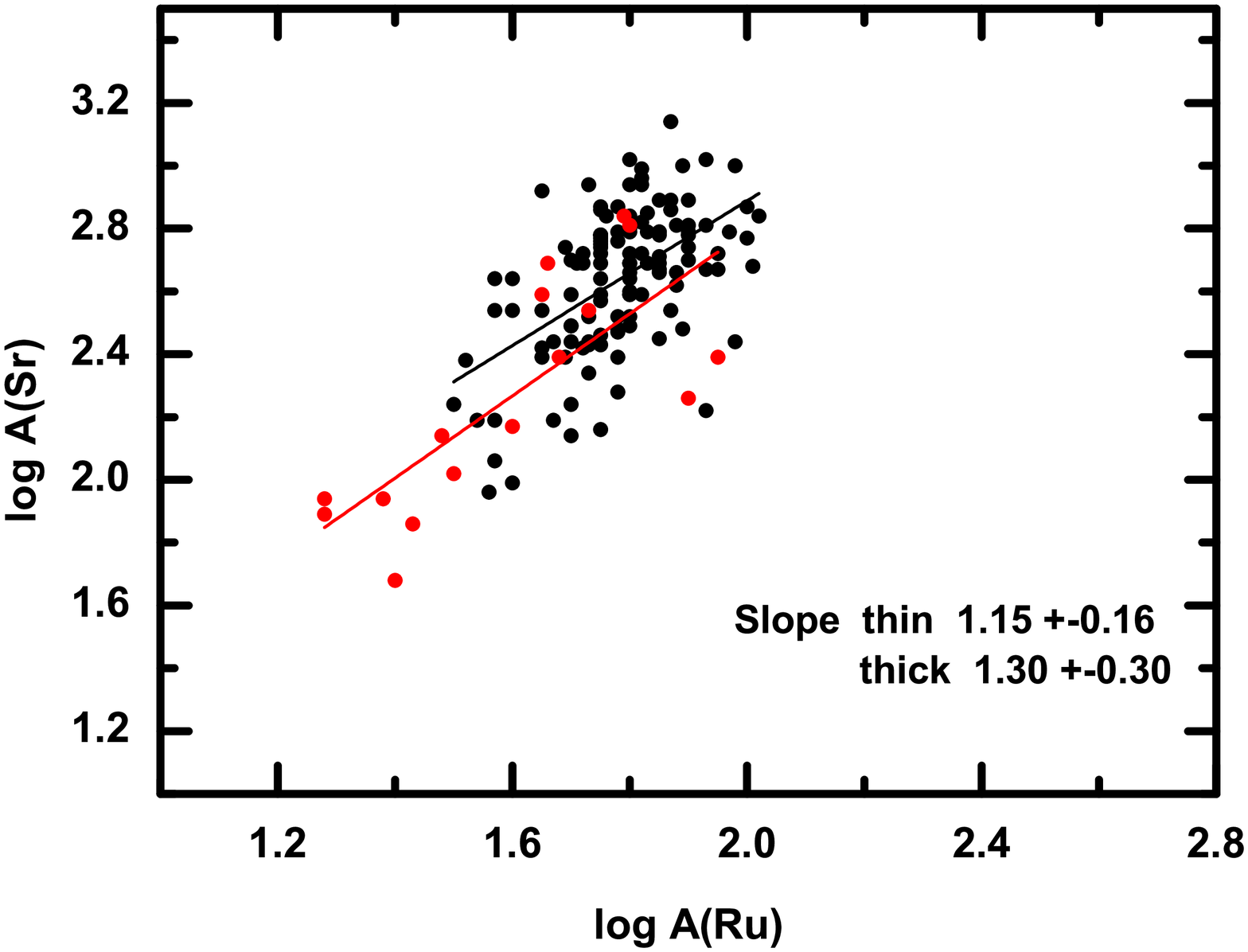}\\
\includegraphics[width=11cm]{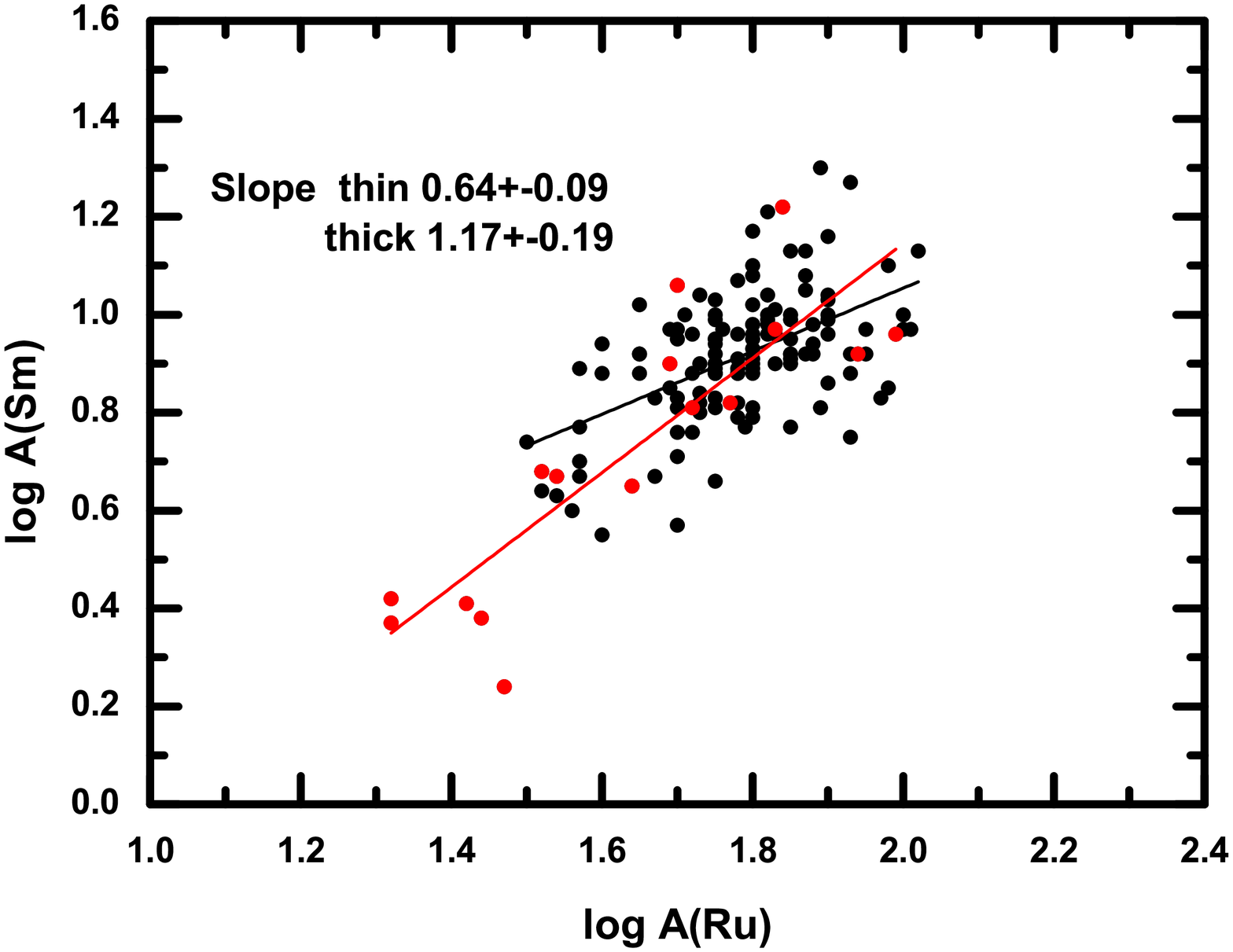}\\
\end{tabular}
\caption{Trends for logA(Sr) vs. logA(Y) and log A(El) where El =  Sr and Sm vs. log A(Ru). Notation is the same as presented in \ref{elmo_n}.}
\label{elmo2_n}
\end{figure*}

The list of stellar parameters and the Mo and Ru abundances is given in Table A1;
The comparison of parameters is presented in Table A2. 
Figs. \ref{elmo1_n}, \ref{elru} and \ref{elmo2_n} A3 illustrate the following correlations: Mo vs. Y, Zr, Sm and Sr; Sr vs. Y; Ru vs. Ba, Eu and Sm.

\onecolumn
\clearpage

\begin{longtable}{lcccccccccc}
\caption{Stellar parameters and abundances of Mo and Ru.}
\label{ncapt}\\
\hline
HD BD    & \Teff, K & \logg  & [Fe/H] & \Vt,km s$^{-1}$ &  [Mo/Fe]& stand deviation & [Ru/Fe] & stand deviation \\ 
\hline  
thin disc&  &  &  &  &  &  &  &    \\
\hline                               
\endfirsthead
\hline
HD/BD    & \Teff, K & \logg  & [Fe/H] & \Vt,km s$^{-1}$ &  [Mo/Fe]& stand deviation & [Ru/Fe] & stand deviation   \\ 
\hline  
thin disc&  &  &  &  &  &  &  &   \\
\endhead
\hline                               
  166   &  5514 &   4.6  &   0.16 &   0.6  &     -0.05 &   0.04  &   --  &    --   \\
  1562  &  5828 &    4   &  -0.32 &   1.2  &      0.18 &   0.04  &  0.29 &   0.08  \\
  1835  &  5790 &   4.5  &   0.13 &   1.1  &       --  &    --   &   --  &    --   \\
  3651  &  5277 &   4.5  &   0.15 &   0.6  &     -0.01 &   0.06  & -0.08 &   0.08  \\
  4256  &  5020 &   4.3  &   0.08 &   1.1  &       --  &    --   &   --  &    --   \\
  4307  &  5889 &    4   &  -0.18 &   1.1  &      0.24 &   0.04  &   --  &    --   \\
  4614  &  5965 &   4.4  &  -0.24 &   1.1  &    $<$ 0.21 &   0.06  &  0.16 &   0.06  \\
  5294  &  5779 &   4.1  &  -0.17 &   1.3  &      0.25 &   0.07  &  0.35 &   0.11  \\
  6660  &  4759 &   4.6  &   0.08 &   1.4  &      0.08 &   0.03  &  0.17 &   0.07  \\
  7590  &  5962 &   4.4  &   -0.1 &   1.4  &    $<$ 0.28 &   0.07  &  0.3  &    --   \\
  7924  &  5165 &   4.4  &  -0.22 &   1.1  &      0.19 &   0.05  &  0.16 &   0.08  \\
  8648  &  5790 &   4.2  &   0.12 &   1.1  &      0.06 &   0.04  &   --  &    --   \\
  9407  &  5666 &   4.45 &   0.05 &   0.8  &     -0.01 &   0.08  &  0.02 &   0.06  \\
  9826  &  6074 &    4   &   0.1  &   1.3  &    $<$ 0.03 & 0.00    &   --  &    --   \\
 10086  &  5696 &   4.3  &   0.13 &   1.2  &      0.02 &   0.04  &  0.02 &    --   \\
 10307  &  5881 &   4.3  &   0.02 &   1.1  &      0.01 &   0.03  &   --  &    --   \\
 10476  &  5242 &   4.3  &  -0.05 &   1.1  &      0.11 &   0.08  &  0.2  &   0.1   \\
 10780  &  5407 &   4.3  &   0.04 &   0.9  &      0.14 &   0.07  & -0.03 &   0.04  \\
 11007  &  5980 &    4   &   -0.2 &   1.1  &      0.18 &   0.07  &  0.18 &   0.09  \\
 11373  &  4783 &   4.65 &   0.08 &    1   &     -0.05 &   0.00    &  0.04 &   0.07  \\
 12846  &  5766 &   4.5  &  -0.24 &   1.2  &      0.37 &    --   &  0.14 &    --   \\
 13507  &  5714 &   4.5  &  -0.02 &   1.1  &      0.1  &   0.00    &  0.1  &   0.06  \\
 14374  &  5449 &   4.3  &  -0.09 &   1.1  &      0.22 &   0.07  &   --  &    --   \\
 16160  &  4829 &   4.6  &  -0.16 &   1.1  &      0.17 &   0.04  &  0.26 &   0.15  \\
 17674  &  5909 &    4   &  -0.14 &   1.1  &      0.07 &   0.07  &  0.28 &   0.01  \\
 17925  &  5225 &   4.3  &  -0.04 &   1.1  &      0.22 &   0.07  &   --  &    --   \\
 18632  &  5104 &   4.4  &   0.06 &   1.4  &      0.16 &   0.06  &   --  &    --   \\
 18803  &  5665 &   4.55 &   0.14 &   0.8  &     -0.11 &   0.07  & -0.11 &   0.03  \\
 19019  &  6063 &    4   &  -0.17 &   1.1  &    $<$ 0.18 &   0.04  &   --  &    --   \\
 19373  &  5963 &   4.2  &   0.06 &   1.1  &    $<$-0.02 &   0.05  & -0.12 &   0.01  \\
 20630  &  5709 &   4.5  &   0.08 &   1.1  &     -0.03 &   0.04  &  0.05 &   0.1   \\
 22049  &  5084 &   4.4  &  -0.15 &   1.1  &      0.18 &   0.07  &   --  &    --   \\
 22484  &  6037 &   4.1  &  -0.03 &   1.1  &    $<$ 0.04 &  0.00    & -0.15 &   0.03  \\
 22556  &  6155 &   4.2  &  -0.17 &   1.1  &       --  &    --   &   --  &    --   \\
 24053  &  5723 &   4.4  &   0.04 &   1.1  &      0.11 &   0.04  &  0.23 &   0.03  \\
 24238  &  4996 &   4.3  &  -0.46 &    1   &      0.27 &   0.04  &  0.41 &   0.07  \\
 24496  &  5536 &   4.3  &  -0.13 &   1.5  &      0.21 &   0.00    &  0.08 &   0.07  \\
 25665  &  4967 &   4.7  &   0.01 &   1.2  &     -0.01 &   0.00   &   --  &    --   \\
 25680  &  5843 &   4.5  &   0.05 &   1.1  &      0.05 &   0.04  & -0.05 &    --   \\
 26923  &  5920 &   4.4  &  -0.03 &    1   &     -0.04 &   0.00   &  0.29 &   0.01  \\
 28005  &  5980 &   4.2  &   0.23 &   1.1  &     -0.08 &   0.11  &  0.00  &   0.08  \\
 28447  &  5639 &    4   &  -0.09 &   1.1  &     -0.03 &   0.07  & -0.06 &    --   \\
 29150  &  5733 &   4.3  &   0.00  &   1.1  &     -0.01 &   0.11  & -0.04 &   0.01  \\
 29310  &  5852 &   4.2  &   0.08 &   1.4  &       --  &    --   &   --  &    --   \\
 29645  &  6009 &    4   &   0.14 &   1.3  &     -0.04 &   0.04  &  0.11 &    --   \\
 30495  &  5820 &   4.4  &  -0.05 &   1.3  &    $<$ 0.09 &   0.01  &   --  &    --   \\
 33632  &  6072 &   4.3  &  -0.24 &   1.1  &    $<$-0.03 &   0.07  &   --  &    --   \\
 34411  &  5890 &   4.2  &   0.1  &   1.1  &     -0.09 &   0.03  &   --  &    --   \\
 37008  &  5016 &   4.4  &  -0.41 &   0.8  &      0.14 &   0.04  &  0.2  &   0.04  \\
 37394  &  5296 &   4.5  &   0.09 &   1.1  &      0.08 &   0.13  &   --  &    --   \\
 38858  &  5776 &   4.3  &  -0.23 &   1.1  &      0.13 &   0.04  &  0.26 &   0.04  \\
 39587  &  5955 &   4.3  &  -0.03 &   1.5  &     -0.09 &   0.07  &  0.16 &   0.04  \\
 40616  &  5881 &    4   &  -0.22 &   1.1  &      0.13 &   0.04  &  0.12 &   0.07  \\
 41330  &  5904 &   4.1  &  -0.18 &   1.2  &      0.19 &   0.00   &   --  &    --   \\
 41593  &  5312 &   4.3  &  -0.04 &   1.1  &      0.19 &   0.05  &  0.26 &   0.1   \\
 42618  &  5787 &   4.5  &  -0.07 &    1   &      0.08 &   0.11  &  0.12 &   0.07  \\
 42807  &  5719 &   4.4  &  -0.03 &   1.1  &      0.06 &   0.07  &  0.13 &   0.07  \\
 43587  &  5927 &   4.1  &  -0.11 &   1.3  &      0.12 &   0.04  &  0.11 &    --   \\
 43856  &  6143 &   4.1  &  -0.19 &   1.1  &       --  &    --   &   --  &    --   \\
 43947  &  6001 &   4.3  &  -0.24 &   1.1  &       --  &    --   &   --  &    --   \\
 45088  &  4959 &   4.3  &  -0.21 &   1.2  &      0.09 &    --   &   --  &    --   \\
 47752  &  4613 &   4.6  &  -0.05 &   0.2  &      0.03 &   0.07  &  0.05 &   0.07  \\
 48682  &  5989 &   4.1  &   0.05 &   1.3  &     -0.02 &   0.06  &  0.05 &   0.00   \\
 50281  &  4712 &   3.9  &   -0.2 &   1.6  &       --  &    --   &   --  &    --   \\
 50692  &  5911 &   4.5  &   -0.1 &   0.9  &      0.21 &   0.04  &  0.00  &    --   \\
 51419  &  5746 &   4.1  &  -0.37 &   1.1  &      0.24 &   0.01  &   --  &    --   \\
 51866  &  4934 &   4.4  &   0.00  &    1   &      0.08 &   0.04  &  0.00  &    --   \\
 53927  &  4860 &   4.64 &  -0.22 &   1.2  &      0.01 &   0.05  &  0.14 &   0.12  \\
 54371  &  5670 &   4.2  &   0.06 &   1.2  &      0.07 &   0.07  &  0.12 &    --   \\
 55575  &  5949 &   4.3  &  -0.31 &   1.1  &      0.09 &   0.07  &  0.26 &   0.05  \\
 58595  &  5707 &   4.3  &  -0.31 &   1.2  &      0.29 &   0.07  &  0.54 &   0.11  \\
 59747  &  5126 &   4.4  &  -0.04 &   1.1  &      0.12 &   0.00   &  0.04 &    --   \\
 61606  &  4956 &   4.4  &  -0.12 &   1.3  &      0.24 &   0.08  &   --  &    --   \\
 62613  &  5541 &   4.4  &   -0.1 &   1.1  &      0.16 &   0.04  &  0.17 &   0.11  \\
 63433  &  5693 &   4.35 &  -0.06 &   1.9  &      0.21 &   0.04  &  0.11 &   0.07  \\
 64468  &  5014 &   4.2  &   0.00  &   1.2  &      0.18 &   0.07  &  0.03 &   0.08  \\
 64815  &  5864 &    4   &  -0.33 &   1.1  &      0.26 &    --   &   --  &    --   \\
 65874  &  5936 &    4   &   0.05 &   1.3  &     -0.11 &   0.02  &   --  &    --   \\
 66573  &  5821 &   4.6  &  -0.53 &   1.1  &      0.32 &   0.01  &  0.57 &   0.06  \\
 68638  &  5430 &   4.4  &  -0.24 &   1.1  &      0.25 &   0.02  &   --  &    --   \\
 70923  &  5986 &   4.2  &   0.06 &   1.1  &     -0.07 &   0.06  &  0.04 &    --   \\
 71148  &  5850 &   4.2  &   0.00  &   1.1  &      0.02 &   0.01  &  0.05 &   0.00   \\
 72760  &  5349 &   4.1  &   0.01 &   1.1  &       --  &    --   &   --  &    --   \\
 72905  &  5884 &   4.4  &  -0.07 &   1.5  &      0.13 &   0.04  &  0.12 &    --   \\
 73344  &  6060 &   4.1  &   0.08 &   1.1  &    $<$-0.02 &   0.11  &  0.07 &    --   \\
 73667  &  4884 &   4.4  &  -0.58 &   0.9  &      0.16 &   0.03  &  0.39 &   0.1   \\
 75732  &  5373 &   4.3  &   0.25 &   1.1  &      0.01 &   0.04  & -0.11 &   0.07  \\
 75767  &  5823 &   4.2  &  -0.01 &   0.9  &     -0.06 &   0.00   &   --  &    --   \\
 76151  &  5776 &   4.4  &   0.05 &   1.1  &     -0.02 &   0.07  &   --  &    --   \\
 79969  &  4825 &   4.4  &  -0.05 &    1   &       --  &    --   &   --  &    --   \\
 82106  &  4827 &   4.1  &  -0.11 &   1.1  &      0.22 &   0.04  &  0.09 &   0.03  \\
 82443  &  5334 &   4.4  &  -0.03 &   1.3  &     -0.01 &   0.09  &   --  &    --   \\
 87883  &  5015 &   4.4  &   0.00  &   1.1  &      0.03 &   0.07  &   --  &    --   \\
 88072  &  5778 &   4.3  &   0.00  &   1.1  &      0.03 &   0.06  & -0.05 &    --   \\
 89251  &  5886 &    4   &  -0.12 &   1.1  &      0.05 &   0.07  &   --  &    --   \\
 89269  &  5674 &   4.4  &  -0.23 &   1.1  &      0.06 &   0.07  &  0.21 &   0.06  \\
 91347  &  5931 &   4.4  &  -0.43 &   1.1  &      0.31 &   0.00   &  0.43 &    --   \\
 94765  &  5077 &   4.4  &  -0.01 &   1.1  &      0.14 &   0.07  &  0.08 &   0.04  \\
 95128  &  5887 &   4.3  &   0.01 &   1.1  &     -0.03 &   0.07  &  0.19 &    --   \\
 97334  &  5869 &   4.4  &   0.06 &   1.2  &      0.00  &   0.11  &   --  &    --   \\
 97658  &  5136 &   4.5  &  -0.32 &   1.2  &      0.12 &   0.02  &  0.07 &   0.03  \\
 98630  &  6060 &    4   &   0.22 &   1.4  &       --  &    --   &   --  &    --   \\
 101177 &  5932 &   4.1  &  -0.16 &   1.1  &    $<$ 0.15 &   0.06  &  0.21 &    --   \\
 102870 &  6055 &    4   &   0.13 &   1.4  &    $<$-0.15 &   0.00   & -0.13 &   0.00   \\
 105631 &  5416 &   4.4  &   0.16 &   1.2  &       --  &    --   &   --  &    --   \\
 107705 &  6040 &   4.2  &   0.06 &   1.4  &    $<$ 0.04 &   0.04  &   --  &    --   \\
 108954 &  6037 &   4.4  &  -0.12 &   1.1  &    $<$ 0.1  &   0.07  &   --  &    --   \\
 109358 &  5897 &   4.2  &  -0.18 &   1.1  &      0.06 &   0.07  &  0.18 &   0.05  \\
 110463 &  4950 &   4.5  &  -0.05 &   1.2  &      0.13 &   0.07  &  0.13 &   0.08  \\
 110833 &  5075 &   4.3  &   0.00  &   1.1  &      0.13 &   0.00   &  0.00  &   0.07  \\
 111395 &  5648 &   4.6  &   0.1  &   0.9  &       --  &    --   &   --  &    --   \\
 112758 &  5203 &   4.2  &  -0.56 &   1.1  &       --  &    --   &   --  &    --   \\
 114710 &  5954 &   4.3  &   0.07 &   1.1  &     -0.06 &   0.05  & -0.04 &    --   \\
 115383 &  6012 &   4.3  &   0.11 &   1.1  &    $<$-0.02 &   0.00   &   --  &    --   \\
 115675 &  4745 &   4.45 &   0.02 &    1   &       --  &    --   &   --  &    --   \\
 116443 &  4976 &   3.9  &  -0.48 &   1.1  &       --  &    --   &   --  &    --   \\
 116956 &  5386 &   4.55 &   0.08 &   1.2  &     -0.03 &   0.04  & -0.03 &   0.07  \\
 117043 &  5610 &   4.5  &   0.21 &   0.4  &     -0.11 &   0.02  & -0.14 &   0.04  \\
 119802 &  4763 &    4   &  -0.05 &   1.1  &      0.06 &   0.04  & -0.05 &    --   \\
 122064 &  4937 &   4.5  &   0.07 &   1.1  &       --  &    --   &   --  &    --   \\
 124642 &  4722 &   4.65 &   0.02 &   1.3  &       --  &    --   &   --  &    --   \\
 125184 &  5695 &   4.3  &   0.31 &   0.7  &     -0.18 &   0.00   & -0.13 &   0.04  \\
 126053 &  5728 &   4.2  &  -0.32 &   1.1  &    $<$ 0.1  &   0.00   &  0.35 &   0.04  \\
 127506 &  4542 &   4.6  &  -0.08 &   1.2  &       --  &    --   &   --  &    --   \\
 128311 &  4960 &   4.4  &   0.03 &   1.3  &      0.1  &   0.07  &  0.02 &   0.07  \\
 130307 &  4990 &   4.3  &  -0.25 &   1.4  &      0.28 &   0.07  &  0.23 &   0.04  \\
 130948 &  5943 &   4.4  &  -0.05 &   1.3  &    $<$ 0.06 &   0.04  &   --  &    --   \\
 131977 &  4683 &   3.7  &  -0.24 &   1.8  &       --  &    --   &   --  &    --   \\
 135599 &  5257 &   4.3  &  -0.12 &    1   &      0.2  &   0.00   &  0.22 &   0.00   \\
 137107 &  6037 &   4.3  &   0.00  &   1.1  &       --  &    --   &   --  &    --   \\
 139777 &  5771 &   4.4  &   0.01 &   1.3  &       --  &    --   &   --  &    --   \\
 139813 &  5408 &   4.5  &   0.00  &   1.2  &      0.08 &   0.00   &   --  &    --   \\
 140538 &  5675 &   4.5  &   0.02 &   0.9  &      0.03 &   0.06  &  0.13 &    --   \\
 141004 &  5884 &   4.1  &  -0.02 &   1.1  &    $<$-0.03 &   0.04  & -0.13 &    --   \\
 141272 &  5311 &   4.4  &  -0.06 &   1.3  &      0.09 &   0.07  &  0.06 &   0.07  \\
 142267 &  5856 &   4.5  &  -0.37 &   1.1  &      0.25 &   0.07  &  0.55 &   0.04  \\
 144287 &  5414 &   4.5  &  -0.15 &   1.1  &      0.16 &   0.04  &  0.18 &   0.04  \\
 145675 &  5406 &   4.5  &   0.32 &   1.1  &       --  &    --   &   --  &    --   \\
 146233 &  5799 &   4.4  &   0.01 &   1.1  &       --  &    --   &   --  &    --   \\
 149661 &  5294 &   4.5  &  -0.04 &   1.1  &      0.07 &   0.07  &  0.01 &   0.04  \\
 149806 &  5352 &   4.55 &   0.25 &   0.4  &       --  &    --   &   --  &    --   \\
 151541 &  5368 &   4.2  &  -0.22 &   1.3  &       --  &    --   &   --  &    --   \\
 153525 &  4810 &   4.7  &  -0.04 &    1   &       --  &    --   &   --  &    --   \\
 154345 &  5503 &   4.3  &  -0.21 &   1.3  &      0.14 &   0.00   &  0.19 &   0.08  \\
 156668 &  4850 &   4.2  &  -0.07 &   1.2  &      0.13 &   0.06  &  0.19 &   0.06  \\
 156985 &  4790 &   4.6  &  -0.18 &    1   &      0.14 &   0.04  &  0.18 &   0.07  \\
 158633 &  5290 &   4.2  &  -0.49 &   1.3  &      0.25 &   0.04  &  0.31 &   0.04  \\
 160346 &  4983 &   4.3  &   -0.1 &   1.1  &      0.18 &   0.07  &  0.15 &    --   \\
 161098 &  5617 &   4.3  &  -0.27 &   1.1  &      0.17 &   0.04  &   --  &    --   \\
 164922 &  5392 &   4.3  &   0.04 &   1.1  &       --  &    --   &   --  &    --   \\
 165173 &  5505 &   4.3  &  -0.05 &   1.1  &      0.04 &   0.09  &   --  &    --   \\
 165341 &  5314 &   4.3  &  -0.08 &   1.1  &      0.16 &   0.07  &  0.13 &    --   \\
 165476 &  5845 &   4.1  &  -0.06 &   1.1  &       --  &    --   &   --  &    --   \\
 165670 &  6178 &    4   &   -0.1 &   1.5  &       --  &    --   &   --  &    --   \\
 165908 &  5925 &   4.1  &   -0.6 &   1.1  &    $<$ 0.28 &   0.07  &  0.45 &    --   \\
 166620 &  5035 &    4   &  -0.22 &    1   &      0.12 &   0.04  &  0.17 &    --   \\
 171314 &  4608 &   4.65 &   0.07 &    1   &       --  &    --   &   --  &    --   \\
 174080 &  4764 &   4.55 &   0.04 &    1   &      0.01 &   0.04  & -0.04 &    --   \\
 175742 &  5030 &   4.5  &  -0.03 &    2   &       --  &    --   &   --  &    --   \\
 176377 &  5901 &   4.4  &  -0.17 &   1.3  &    $<$ 0.2  &   0.00   &  0.22 &    --   \\
 176841 &  5841 &   4.3  &   0.23 &   1.1  &    $<$ -0.1 &   0.07  & -0.08 &    --   \\
 178428 &  5695 &   4.4  &   0.14 &    1   &     -0.11 &   0.00   &  0.01 &    --   \\
 180161 &  5473 &   4.5  &   0.18 &   1.1  &      0.02 &   0.09  & -0.13 &    --   \\
 182488 &  5435 &   4.4  &   0.07 &   1.1  &      0.06 &   0.07  & -0.07 &    --   \\
 183341 &  5911 &   4.3  &  -0.01 &   1.3  &       --  &    --   &   --  &    --   \\
 184385 &  5536 &   4.45 &   0.12 &   0.9  &     -0.07 &   0.04  & -0.12 &   0.05  \\
 185144 &  5271 &   4.2  &  -0.33 &   1.1  &      0.16 &   0.00   &  0.1  &   0.04  \\
 185414 &  5818 &   4.3  &  -0.04 &   1.1  &       --  &    --   &   --  &    --   \\
 186408 &  5803 &   4.2  &   0.09 &   1.1  &     -0.01 &   0.07  &  0.06 &    --   \\
 186427 &  5752 &   4.2  &   0.02 &   1.1  &     -0.01 &   0.01  &   --  &    --   \\
 187897 &  5887 &   4.3  &   0.08 &   1.1  &      0.00  &   0.07  &  0.02 &   0.00   \\
 189087 &  5341 &   4.4  &  -0.12 &   1.1  &       --  &    --   &   --  &    --   \\
 189733 &  5076 &   4.4  &  -0.03 &   1.5  &     -0.01 &   0.07  &  0.16 &   0.04  \\
 190007 &  4724 &   4.5  &   0.16 &   0.8  &      0.1  &   0.04  & -0.11 &    --   \\
 190406 &  5905 &   4.3  &   0.05 &    1   &     -0.02 &   0.03  &   --  &    --   \\
 190470 &  5130 &   4.3  &   0.11 &    1   &      0.02 &   0.07  & -0.03 &   0.03  \\
 190771 &  5766 &   4.3  &   0.13 &   1.5  &      0.05 &   0.07  &   --  &    --   \\
 191533 &  6167 &   3.8  &   -0.1 &   1.5  &       --  &    --   &   --  &    --   \\
 191785 &  5205 &   4.2  &  -0.12 &   1.2  &      0.14 &   0.08  &  0.15 &   0.04  \\
 195005 &  6075 &   4.2  &  -0.06 &   1.3  &       --  &    --   &   --  &    --   \\
 195104 &  6103 &   4.3  &  -0.19 &   1.1  &       --  &    --   &   --  &    --   \\
 197076 &  5821 &   4.3  &  -0.17 &   1.2  &      0.2  &    --   &  0.12 &    --   \\
 199960 &  5878 &   4.2  &   0.23 &   1.1  &     -0.13 &   0.04  & -0.18 &    --   \\
 200560 &  5039 &   4.4  &   0.06 &   1.1  &      0.12 &   0.00   &  0.04 &    --   \\
 202108 &  5712 &   4.2  &  -0.21 &   1.1  &      0.17 &   0.04  &  0.16 &    --   \\
 202575 &  4667 &   4.6  &  -0.03 &   0.5  &     -0.01 &   0.05  &  0.00  &   0.04  \\
 203235 &  6071 &   4.1  &   0.05 &   1.3  &       --  &    --   &   --  &    --   \\
 205702 &  6020 &   4.2  &   0.01 &   1.1  &       --  &    --   &   --  &    --   \\
 206860 &  5927 &   4.6  &  -0.07 &   1.8  &       --  &    --   &   --  &    --   \\
 208038 &  4982 &   4.4  &  -0.08 &    1   &       --  &    --   &   --  &    --   \\
 208313 &  5055 &   4.3  &  -0.05 &    1   &      0.06 &   0.04  &  0.1  &   0.03  \\
 208906 &  5965 &   4.2  &   -0.8 &   1.7  &       --  &    --   &   --  &    --   \\
 210667 &  5461 &   4.5  &   0.15 &   0.9  &      0.08 &   0.07  & -0.08 &   0.04  \\
 210752 &  6014 &   4.6  &  -0.53 &   1.1  &    $<$ 0.26 &   0.00   &   --  &    --   \\
 211472 &  5319 &   4.4  &  -0.04 &   1.1  &      0.15 &   0.04  &  0.04 &   0.05  \\
 214683 &  4747 &   4.6  &  -0.46 &   1.2  &      0.11 &   0.04  &  0.28 &   0.05  \\
 216259 &  4833 &   4.6  &  -0.55 &   0.5  &       --  &    --   &   --  &    --   \\
 216520 &  5119 &   4.4  &  -0.17 &   1.4  &       --  &    --   &   --  &    --   \\
 217014 &  5763 &   4.3  &   0.17 &   1.1  &     -0.06 &   0.11  & -0.05 &   0.06  \\
 217813 &  5845 &   4.3  &   0.03 &   1.5  &      0.1  &    --   &  0.02 &    --   \\
 218868 &  5547 &   4.45 &   0.21 &   0.4  &       --  &    --   &   --  &    --   \\
 219538 &  5078 &   4.5  &  -0.04 &   1.1  &       --  &    --   &   --  &    --   \\
 219623 &  5949 &   4.2  &   0.04 &   1.2  &      0.09 &    --   & -0.04 &    --   \\
 220140 &  5144 &   4.6  &  -0.03 &   2.4  &       --  &    --   &   --  &    --   \\
 220182 &  5364 &   4.5  &  -0.03 &   1.2  &      0.08 &   0.04  &  0.13 &    --   \\
 220221 &  4868 &   4.5  &   0.16 &   0.5  &      0.05 &   0.04  & -0.04 &   0.06  \\
 221851 &  5184 &   4.4  &  -0.09 &    1   &       --  &    --   &   --  &    --   \\
 222143 &  5823 &   4.45 &   0.15 &   1.1  &     -0.04 &   0.05  & -0.05 &   0.07  \\
 224465 &  5745 &   4.5  &   0.08 &   0.8  &     -0.07 &   0.04  & -0.08 &    --   \\
 263175 &  4734 &   4.5  &  -0.16 &   0.5  &      0.08 &   0.06  &  0.19 &   0.04  \\
BD12063 &  4859 &   4.4  &  -0.22 &   0.6  &     -0.07 &   0.04  &  0.04 &   0.04  \\
BD124499&  4678 &   4.7  &   0.00  &   0.5  &       --  &    --   &   --  &    --   \\
\hline  
thick disc&  &  &  &  &  &  &  &  \\
\hline                               
  245   &  5400  &  3.4  &  -0.84  &  0.7   &     0.3  &   0.04 &    -- &     --   \\
  3765  &  5079  &  4.3  &   0.01  &  1.1   &     0.09 &   0.02 &   0.07&    0.08  \\
  5351  &  4378  &  4.6  &  -0.21  &  0.5   &      --  &    --  &    -- &     --   \\
  6582  &  5350  &  4.5  &  -0.83  &  0.4   &     0.37 &   0.01 &   0.4 &    0.03  \\
 13783  &  5350  &  4.1  &  -0.75  &  1.1   &     0.33 &    --  &   0.32&    0.04  \\
 18757  &  5741  &  4.3  &  -0.25  &   1    &     0.21 &   0.04 &   0.22&    0.05  \\
 22879  &  5825  &  4.42 &  -0.91  &  0.9   &   $<$ 0.4  &   0.00  &   0.63&    0.04  \\
 65583  &  5373  &  4.6  &  -0.67  &  0.7   &     0.38 &   0.04 &   0.44&    0.04  \\
 76932  &  5840  &   4   &  -0.95  &   1    &   $<$ 0.33 &   0.07 &    -- &     --   \\
 106516 &  6165  &  4.4  &  -0.72  &  1.1   &     0.35 &    --  &    -- &     --   \\
 110897 &  5925  &  4.2  &  -0.45  &  1.1   &     0.21 &   0.04 &   0.34&    0.07  \\
 135204 &  5413  &   4   &  -0.16  &  1.1   &     0.19 &   0.00  &   0.1 &    0.07  \\
 152391 &  5495  &  4.3  &  -0.08  &  1.3   &     0.16 &   0.07 &   0.03&    0.07  \\
 157089 &  5785  &   4   &  -0.56  &   1    &     0.27 &   0.04 &    -- &     --   \\
 157214 &  5820  &  4.5  &  -0.29  &   1    &     0.27 &   0.1  &   0.53&    0.07  \\
 159062 &  5414  &  4.3  &   -0.4  &   1    &     0.38 &   0.03 &   0.42&    0.08  \\
 165401 &  5877  &  4.3  &  -0.36  &  1.1   &     0.27 &   0.04 &   0.55&     --   \\
 190360 &  5606  &  4.4  &   0.12  &  1.1   &     0.06 &   0.14 &  -0.03&     --   \\
 201889 &  5600  &  4.1  &  -0.85  &  1.2   &     0.63 &   0.07 &   0.52&     --   \\
 201891 &  5850  &  4.4  &  -0.96  &   1    &     0.47 &   0.04 &   0.65&     --   \\
 204521 &  5809  &  4.6  &  -0.66  &  1.1   &     0.47 &   0.11 &   0.45&     --   \\
\hline 
 Hercules stream   &  &  &  &  &  &  &  &    \\
\hline  
 13403   & 5724  &   4  &   -0.31  &  1.1&        0.19 &   0.03 &   0.21 &   0.07  \\
 19308   & 5844  &  4.3 &    0.08  &  1.1&       -0.08 &   0.11 &  -0.03 &   0.05  \\
 23050   & 5929  &  4.4 &   -0.36  &  1.1&        0.34&   0.07 &   0.31 &    --   \\
 30562   & 5859  &   4  &    0.18  &  1.1&         --  &    --  &    --  &    --   \\
 64606   & 5250  &  4.2 &   -0.91  &  0.8&        0.40&   0.02 &   0.66 &   0.07  \\
 68017   & 5651  &  4.2 &   -0.42  &  1.1&        0.25&   0.00  &   0.32 &    --   \\
 81809   & 5782  &   4  &   -0.28  &  1.3&      $<$ 0.31 &   0.07 &   0.28 &    --   \\
 107213  & 6156  &  4.1 &    0.07  &  1.6&      $<$0.01 &   0.00  &    --  &    --   \\
 139323  & 5204  &  4.6 &    0.19  &  0.7&         --  &    --  &    --  &    --   \\
 139341  & 5242  &  4.6 &    0.21  &  0.9&         --  &    --  &    --  &    --   \\
 144579  & 5294  &  4.1 &    -0.7  &  1.3&        0.40&   0.04 &   0.45 &   0.07  \\
 159222  & 5834  &  4.3 &    0.06  &  1.2&        0.07 &   0.07 &   0.01 &   0.04  \\
 159909  & 5749  &  4.1 &    0.06  &  1.1&        -0.08 &   0.07 &    --  &    --   \\
 215704  & 5418  &  4.2 &    0.07  &  1.1&         --  &    --  &    --  &    --   \\
 218209  & 5705  &  4.5 &   -0.43  &   1 &        0.31&    --  &   0.38 &    --   \\
 221354  & 5242  &  4.1 &   -0.06  &  1.2&        0.13&   0.08 &  -0.01 &    --   \\
\hline 
 nonclassified   &  &  &  &  &  &  &  &   \\
\hline
  4628  &  4905 &   4.6  &  -0.36 &   0.5 &      0.21 &   0.04 &   0.21  &   --   \\
  4635  &  5103 &   4.4  &   0.07 &   0.8 &      0.08 &   0.04 &   0.13  &  0.07  \\
 10145  &  5673 &   4.4  &  -0.01 &   1.1 &       --  &    --  &    --   &   --   \\
 12051  &  5458 &   4.55 &   0.24 &   0.5 &     -0.12 &   0.07 &  -0.24  &   --   \\
 13974  &  5590 &   3.8  &  -0.49 &   1.1 &       --  &    --  &    --   &   --   \\
 17660  &  4713 &   4.75 &   0.17 &   1.3 &       --  &    --  &    --   &   --   \\
 20165  &  5145 &   4.4  &  -0.08 &   1.1 &      0.20 &   0.07 &   0.13  &  0.07  \\
 24206  &  5633 &   4.5  &  -0.08 &   1.1 &      0.15 &   0.07 &   0.06  &  0.04  \\
 32147  &  4945 &   4.4  &   0.13 &   1.1 &       --  &    --  &    --   &   --   \\
 45067  &  6058 &    4   &  -0.02 &   1.2 &       --  &    --  &    --   &   --   \\
 84035  &  4808 &   4.8  &   0.25 &   0.5 &       --  &    --  &    --   &   --   \\
 86728  &  5725 &   4.3  &   0.22 &   0.9 &       --  &    --  &    --   &   --   \\
 90875  &  4788 &   4.5  &   0.24 &   0.5 &       --  &    --  &    --   &   --   \\
 117176 &  5611 &    4   &  -0.03 &    1  &      0.05 &   0.03 &  -0.01  &  0.00  \\
 117635 &  5230 &   4.3  &  -0.46 &   0.7 &      0.33 &   0.07 &    --   &   --   \\
 154931 &  5910 &    4   &   -0.1 &   1.1 &       --  &    --  &    --   &   --   \\
 159482 &  5620 &   4.1  &  -0.89 &    1  &       --  &    --  &    --   &   --   \\
 168009 &  5826 &   4.1  &  -0.01 &   1.1 &      0.11 &   0.07 &  -0.03  &   --   \\
 173701 &  5423 &   4.4  &   0.18 &   1.1 &      0.04 &   0.07 &  -0.03  &   --   \\
 182736 &  5430 &   3.7  &  -0.06 &    1  &       --  &    --  &    --   &   --   \\
 184499 &  5750 &    4   &  -0.64 &   1.5 &     $<$0.34 &   0.04 &   0.69  &   --   \\
 184768 &  5713 &   4.2  &  -0.07 &   1.1 &       --  &    --  &    --   &   --   \\
 186104 &  5753 &   4.2  &   0.05 &   1.1 &       --  &    --  &    --   &   --   \\
 215065 &  5726 &    4   &  -0.43 &   1.1 &      0.30 &   0.07 &   0.38  &  0.07  \\
 219134 &  4900 &   4.2  &   0.05 &   0.8 &      0.12 &   0.00 &    --   &   --   \\
 219396 &  5733 &    4   &   -0.1 &   1.2 &      0.10 &   0.04 &   0.20  &   --   \\
 224930 &  5300 &   4.1  &  -0.91 &   0.7 &      0.26 &   0.04 &   0.26  &   --   \\
\hline
\end{longtable}

\begin{table*}
\begin{center}
\caption{Parameters of our target stars and their comparison with with those reported in \protect\citet{adibekyan:14, nissen:11, feltzing:07, takeda:07} for common stars.}
\label{comp_edd}
\begin{tabular}{lcccccccccc}
\hline
HD & \Teff, K & \logg & [Fe/H] &  HD & \Teff, K  &  \logg  & [Fe/H]& $\Delta$ \Teff, K& $\Delta$ \logg 
	   & $\Delta$ [Fe/H]   \\
\hline
     & our    &    &     &  \citet{adibekyan:14} &       &     &     &     &      &      \\ 
\hline 
	4307	&	5889	&	4	&	-0.18	&		4307	&	5812	&	4.1	&	-0.23	&	77	&	-0.1	&	0.05	\\
	14374	&	5449	&	4.3	&	-0.09	&		14374	&	5425	&	4.48	&	-0.04	&	24	&	-0.18	&	-0.05	\\
	22879	&	5972	&	4.5	&	-0.77	&		22879	&	5884	&	4.52	&	-0.82	&	88	&	-0.02	&	0.05	\\
	38858	&	5776	&	4.3	&	-0.23	&		38858	&	5733	&	4.51	&	-0.22	&	43	&	-0.21	&	-0.01	\\
	125184	&	5695	&	4.3	&	0.31	&		125184	&	5680	&	4.1	&	0.27	&	15	&	0.2	&	0.04	\\
	146233	&	5799	&	4.4	&	0.01	&		146233	&	5818	&	4.45	&	0.04	&	-19	&	-0.05	&	-0.03	\\
	161098	&	5617	&	4.3	&	-0.27	&		161098	&	5560	&	4.46	&	-0.27	&	57	&	-0.16	&	0	\\
	199960	&	5878	&	4.2	&	0.23	&		199960	&	5973	&	4.39	&	0.28	&	-95	&	-0.19	&	-0.05	\\
	210752	&	6014	&	4.6	&	-0.53	&		210752	&	5951	&	4.53	&	-0.58	&	63	&	0.07	&	0.05	\\
\hline
    &  our    &    &     &    \citet{nissen:11}&      &     &     &     &      &      \\ 
\hline
	22879	&	5972	&	4.5	&	-0.77	&		22879	&	5759	&	4.25	&	-0.85	&	213	&	0.25	&	0.08	\\
	76932	&	5840	&	4	&	-0.95	&		76932	&	5877	&	4.13	&	-0.87	&	-37	&	-0.13	&	-0.08	\\
	106516	&	6165	&	4.4	&	-0.72	&		106516	&	6196	&	4.42	&	-0.68	&	-31	&	-0.02	&	-0.04	\\
	159482	&	5620	&	4.1	&	-0.89	&		159482	&	5737	&	4.31	&	-0.73	&	-117	&	-0.21	&	-0.16	\\
\hline
      &  our    &    &     &    \citet{feltzing:07} &     &     &     &     &      &      \\ 
\hline
	22879	&	5972	&	4.5	&	-0.77	&		22879	&	5920	&	4.33	&	-0.84	&	52	&	0.17	&	0.07	\\
	30495	&	5820	&	4.4	&	-0.05	&		30495	&	5850	&	4.5	&	0.05	&	-30	&	-0.1	&	-0.1	\\
	76932	&	5840	&	4	&	-0.95	&		76932	&	5875	&	4.1	&	-0.91	&	-35	&	-0.1	&	-0.04	\\
	152391	&	5495	&	4.3	&	-0.08	&		152391	&	5470	&	4.55	&	-0.02	&	25	&	-0.25	&	-0.06	\\
	157089	&	5785	&	4	&	-0.56	&		157089	&	5830	&	4.06	&	-0.57	&	-45	&	-0.06	&	0.01	\\
	165401	&	5877	&	4.3	&	-0.36	&		165401	&	5720	&	4.35	&	-0.46	&	157	&	-0.05	&	0.1	\\
	176377	&	5901	&	4.4	&	-0.17	&		176377	&	5810	&	4.4	&	-0.28	&	91	&	0	&	0.11	\\
	190360	&	5606	&	4.4	&	0.12	&		190360	&	5490	&	4.23	&	0.25	&	116	&	0.17	&	-0.13	\\
	199960	&	5878	&	4.2	&	0.23	&		199960	&	5940	&	4.26	&	0.27	&	-62	&	-0.06	&	-0.04	\\
	217014	&	5763	&	4.3	&	0.17	&		217014	&	5789	&	4.34	&	0.2	&	-26	&	-0.04	&	-0.03	\\
\hline
     &  our    &    &     &    \citet{takeda:07} &     &     &     &     &      &      \\ 
\hline
	4307	&	5889	&	4	&	-0.18	&		4307	&	5648.1	&	3.747	&	-0.289	&	240.9	&	0.253	&	0.109	\\
	4614	&	5965	&	4.4	&	-0.24	&		4614	&	5915.4	&	4.462	&	-0.214	&	49.6	&	-0.062	&	-0.026	\\
	4628	&	4905	&	4.6	&	-0.36	&		4628	&	5009.3	&	4.62	&	-0.243	&	-104.3	&	-0.02	&	-0.117	\\
	6582	&	5240	&	4.3	&	-0.94	&		6582	&	5330.7	&	4.539	&	-0.811	&	-90.7	&	-0.239	&	-0.129	\\
	10307	&	5881	&	4.3	&	0.02	&		10307	&	5890.9	&	4.36	&	0.058	&	-9.9	&	-0.06	&	-0.038	\\
	10476	&	5242	&	4.3	&	-0.05	&		10476	&	5196.3	&	4.504	&	-0.011	&	45.7	&	-0.204	&	-0.039	\\
	10780	&	5407	&	4.3	&	0.04	&		10780	&	5427.1	&	4.632	&	0.1	&	-20.1	&	-0.332	&	-0.06	\\
	17925	&	5225	&	4.3	&	-0.04	&		17925	&	5235.2	&	4.669	&	0.133	&	-10.2	&	-0.369	&	-0.173	\\
	18803	&	5665	&	4.55	&	0.14	&		18803	&	5665.7	&	4.455	&	0.146	&	-0.7	&	0.095	&	-0.006	\\
	20630	&	5709	&	4.5	&	0.08	&		20630	&	5768.6	&	4.544	&	0.107	&	-59.6	&	-0.044	&	-0.027	\\
	30562	&	5859	&	4	&	0.18	&		30562	&	5908.3	&	4.084	&	0.232	&	-49.3	&	-0.084	&	-0.052	\\
	34411	&	5890	&	4.2	&	0.1	&		34411	&	5888.6	&	4.232	&	0.107	&	1.4	&	-0.032	&	-0.007	\\
	81809	&	5782	&	4	&	-0.28	&		81809	&	5619.6	&	4.018	&	-0.345	&	162.4	&	-0.018	&	0.065	\\
	86728	&	5725	&	4.3	&	0.22	&		86728	&	5837.9	&	4.421	&	0.268	&	-112.9	&	-0.121	&	-0.048	\\
	110897	&	5925	&	4.2	&	-0.45	&		110897	&	5841.6	&	4.325	&	-0.503	&	83.4	&	-0.125	&	0.053	\\
	111395	&	5648	&	4.6	&	0.1	&		111395	&	5631.7	&	4.485	&	0.105	&	16.3	&	0.115	&	-0.005	\\
	115383	&	6012	&	4.3	&	0.11	&		115383	&	6119.7	&	4.251	&	0.214	&	-107.7	&	0.049	&	-0.104	\\
	117176	&	5611	&	4	&	-0.03	&		117176	&	5466.4	&	3.799	&	-0.112	&	144.6	&	0.201	&	0.082	\\
	125184	&	5695	&	4.3	&	0.31	&		125184	&	5629.6	&	4.015	&	0.247	&	65.4	&	0.285	&	0.063	\\
	141004	&	5884	&	4.1	&	-0.02	&		141004	&	5877	&	4.113	&	-0.008	&	7	&	-0.013	&	-0.012	\\
	149661	&	5294	&	4.5	&	-0.04	&		149661	&	5288.6	&	4.607	&	0.053	&	5.4	&	-0.107	&	-0.093	\\
	157214	&	5820	&	4.5	&	-0.29	&		157214	&	5693.2	&	4.214	&	-0.369	&	126.8	&	0.286	&	0.079	\\
	165908	&	5925	&	4.1	&	-0.6	&		165908	&	6183.1	&	4.347	&	-0.456	&	-258.1	&	-0.247	&	-0.144	\\
	178428	&	5695	&	4.4	&	0.14	&		178428	&	5660.3	&	4.189	&	0.162	&	34.7	&	0.211	&	-0.022	\\
	182488	&	5435	&	4.4	&	0.07	&		182488	&	5417.1	&	4.578	&	0.214	&	17.9	&	-0.178	&	-0.144	\\
	190406	&	5905	&	4.3	&	0.05	&		190406	&	5944.3	&	4.396	&	0.056	&	-39.3	&	-0.096	&	-0.006	\\
	197076	&	5821	&	4.3	&	-0.17	&		197076	&	5804.7	&	4.405	&	-0.075	&	16.3	&	-0.105	&	-0.095	\\
	199960	&	5878	&	4.2	&	0.23	&		199960	&	5924	&	4.26	&	0.275	&	-46	&	-0.06	&	-0.045	\\
	217014	&	5763	&	4.3	&	0.17	&		217014	&	5779.1	&	4.298	&	0.203	&	-16.1	&	0.002	&	-0.033	\\
	219623	&	5949	&	4.2	&	0.04	&		219623	&	6103	&	4.185	&	0.049	&	-154	&	0.015	&	-0.009	\\
	224930	&	5300	&	4.1	&	-0.91	&		224930	&	5680.5	&	4.863	&	-0.522	&	-380.5	&	-0.763	&	-0.388	\\
\hline                                                                                         
\end{tabular} 
\end{center}  
\end{table*}


\label{lastpage}

\bsp

\end{document}